\documentclass[fleqn,usenatbib]{mnras}

\usepackage{newtxtext,newtxmath}

\usepackage[T1]{fontenc}
\usepackage{ae,aecompl}

\usepackage{graphicx}	
\usepackage{amsmath}	
\usepackage[normalem]{ulem}
\usepackage{pdflscape}
\usepackage[dvipsnames]{xcolor}
\usepackage{multirow}

\newcommand{\hMsun}{h^{-1}\mathrm{M_\odot}}
\newcommand{\hMpc}{h^{-1}\mathrm{Mpc}}
\newcommand{\invhMpc}{h\mathrm{Mpc}^{-1}}
\newcommand{\hGpc}{h^{-1}\mathrm{Gpc}}

\newcommand{\kms}{\mathrm{km\,s^{-1}}}

\newcommand{\sqdeg}{\mathrm{deg}^2}

\newcommand{\vect}[1]{\boldsymbol{#1}}
\newcommand{\fsig}{f\sigma_8}
\newcommand{\apar}{\alpha_\parallel}
\newcommand{\aperp}{\alpha_\bot}
\newcommand{\zeff}{z_\mathrm{eff}}
\newcommand{\LCDM}{$\Lambda$CDM}
\newcommand{\dM}{D_\mathrm{M}}
\newcommand{\dH}{D_\mathrm{H}}
\newcommand{\rdrag}{r_\mathrm{drag}}

\newcommand{\Neveux}{RegPT}
\newcommand{\Hou}{\textsc{respresso}}

\newcommand{\highlight}[1]{#1}

\title[eBOSS QSO Mock Challenge]{The Completed SDSS-IV Extended Baryon Oscillation Spectroscopic
Survey: N-body Mock Challenge for the Quasar Sample}

\author[A. Smith]{\parbox{\textwidth}{
Alex Smith\thanks{Email: alexander.smith@cea.fr}$^{1}$,
Etienne Burtin$^{1}$,
Jiamin Hou$^{2}$,
Richard Neveux$^{1}$,
Ashley J. Ross$^{3}$,
Shadab Alam$^{4}$,
Jonathan Brinkmann$^{5}$,
Kyle S. Dawson$^{6}$,
Salman Habib$^{7,8}$,
Katrin Heitmann$^{7}$,
Jean-Paul Kneib$^{9}$,
Brad W. Lyke$^{10}$,
H\'elion du Mas des Bourboux$^{6}$,
Eva-Maria Mueller$^{11}$,
Adam D. Myers$^{10}$,
Will J. Percival$^{12,13,14}$,
Graziano Rossi$^{15}$,
Donald P. Schneider$^{16,17}$,
Pauline Zarrouk$^{18,1}$,
Gong-Bo Zhao$^{19,20}$
 } \vspace*{4pt} \\ 
\scriptsize $^{1}$ IRFU, CEA, Universit\'e Paris-Saclay, F-91191 Gif-sur-Yvette, France\vspace*{-2pt} \\ 
\scriptsize $^{2}$ Max-Planck-Institut f\"ur Extraterrestrische Physik, Postfach 1312, Giessenbachstr., 85748 Garching bei M\"unchen, Germany\vspace*{-2pt} \\ 
\scriptsize $^{3}$ Center for Cosmology and Astro-Particle Physics, Ohio State University, Columbus, Ohio, USA\vspace*{-2pt} \\ 
\scriptsize $^{4}$ Institute for Astronomy, University of Edinburgh, Royal Observatory, Edinburgh, EH9 3HJ, UK\vspace*{-2pt} \\ 
\scriptsize $^{5}$ Apache Point Observatory and New Mexico State University, P.O. Box 59, Sunspot, NM 88349, USA\vspace*{-2pt} \\ 
\scriptsize $^{6}$ Department Physics and Astronomy, University of Utah, 115 S 1400 E, Salt Lake City, UT 84112, USA\vspace*{-2pt} \\ 
\scriptsize $^{7}$ High Energy Physics Division, Argonne National Laboratory, Lemont, IL 60439, USA\vspace*{-2pt} \\ 
\scriptsize $^{8}$ Computational Science Division, Argonne National Laboratory, Lemont, IL 60439, USA\vspace*{-2pt} \\ 
\scriptsize $^{9}$ Institute of Physics, Laboratory of Astrophysics, \'Ecole Polytechnique F\'ed\'erale de Lausanne (EPFL), Observatoire de Sauverny, 1290 Versoix, Switzerland\vspace*{-2pt} \\ 
\scriptsize $^{10}$ Department of Physics and Astronomy, University of Wyoming, Laramie, WY 82071, USA\vspace*{-2pt} \\ 
\scriptsize $^{11}$ Sub-department of Astrophysics, Department of Physics, University of Oxford, Denys Wilkinson Building, Keble Road, Oxford OX1 3RH\vspace*{-2pt} \\ 
\scriptsize $^{12}$ Waterloo Centre for Astrophysics, Department of Physics and Astronomy, University of Waterloo, Waterloo, ON N2L 3G1, Canada\vspace*{-2pt} \\ 
\scriptsize $^{13}$ Department of Physics and Astronomy, University of Waterloo, Waterloo, ON N2L 3G1, Canada\vspace*{-2pt} \\ 
\scriptsize $^{14}$ Perimeter Institute for Theoretical Physics, 31 Caroline St. North, Waterloo, ON N2L 2Y5, Canada\vspace*{-2pt} \\ 
\scriptsize $^{15}$ Department of Physics and Astronomy, Sejong University, Seoul 143-747, Korea\vspace*{-2pt} \\ 
\scriptsize $^{16}$ Department of Astronomy and Astrophysics, The Pennsylvania State University, University Park, PA 16802, USA\vspace*{-2pt} \\ 
\scriptsize $^{17}$ Institute for Gravitation and the Cosmos, The Pennsylvania State University, University Park, PA 16802, USA\vspace*{-2pt} \\ 
\scriptsize $^{18}$ Institute for Computational Cosmology, Dept. of Physics, University of Durham, South Road, Durham DH1 3LE, United Kingdom\vspace*{-2pt} \\ 
\scriptsize $^{19}$ National Astronomy Observatories, Chinese Academy of Science, Beijing, 100101, P.R. China\vspace*{-2pt} \\ 
\scriptsize $^{20}$ Institute of Cosmology \& Gravitation, Dennis Sciama Building, University of Portsmouth, Portsmouth, PO1 3FX, UK\vspace*{-2pt} \\ 
}

\date{Accepted XXX. Received YYY; in original form ZZZ}

\pubyear{2020}

\begin{document}
\label{firstpage}
\pagerange{\pageref{firstpage}--\pageref{lastpage}}
\maketitle

\begin{abstract}
The growth rate and expansion history of the Universe can be measured from
large galaxy redshift surveys using the Alcock-Paczynski effect.
We validate the Redshift Space Distortion models used in the final analysis of the 
Sloan Digital Sky Survey (SDSS) extended Baryon Oscillation Spectroscopic Survey (eBOSS) 
Data Release 16 quasar clustering sample, in configuration and Fourier space, 
using a series of HOD mock catalogues generated using the OuterRim N-body simulation. 
We test three models on a series of non-blind mocks, in the OuterRim cosmology, and blind mocks, which have been
rescaled to new cosmologies, and investigate the effects of redshift smearing and catastrophic redshifts.
We find that for the non-blind mocks, the models are able to recover $\fsig$ to within 3\% and $\apar$ and $\aperp$
to within 1\%. The scatter in the measurements is larger for the blind mocks, due to the assumption of
an incorrect fiducial cosmology.
From this mock challenge, we find that all three models perform well, with similar systematic errors on 
$\fsig$, $\apar$ and $\aperp$ at the level of $\sigma_{\fsig}=0.013$, $\sigma_{\apar}=0.012$ 
and $\sigma_{\aperp}=0.008$. The systematic error on the combined consensus is $\sigma_{\fsig}=0.011$, 
$\sigma_{\apar}=0.008$ and $\sigma_{\aperp}=0.005$, which is used in the final DR16 analysis.
For BAO fits in configuration and Fourier space, we take conservative systematic errors of
$\sigma_{\apar}=0.010$ and $\sigma_{\aperp}=0.007$.
\end{abstract}





\section{Introduction}

The \LCDM\ model of cosmology has been extremely successful in describing
the expansion history and formation of structure in our 
Universe. Measurements covering a wide range of redshifts, from
the temperature fluctuations in the cosmic
microwave background at $z\sim 1100$ \citep{Planck2018}, to galaxy
clustering measurements at late times
\citep[e.g.][]{Eisenstein2005,Alam2017,Abbott2018}, 
are all remarkably consistent with a flat, \LCDM\ Universe. In the standard \LCDM\ model,
gravity follows general relativity, 
dark matter is composed of a collisionless, cold dark matter (CDM), 
and dark energy is described by the cosmological constant, $\Lambda$.
However, this dark energy, which makes up the largest component of the energy 
density ($\approx 70\%$) of the Universe \citep{Planck2018} and is responsible 
for driving the present day accelerated expansion, is poorly understood. 
Constraints can be placed on theories of dark energy and modifications to
general relativity by measuring the expansion history and growth rate of 
structures from surveys of 
galaxies in large cosmological volumes. Reaching the high precision
measurements that are needed to tighten current constraints requires
surveys covering even larger volumes.

One of the primary measurements that is taken to
constrain the nature of dark energy is of the baryon acoustic oscillation
\citep[BAO; e.g.][]{Cole2005, Eisenstein2005}.
In the early Universe, acoustic waves in the hot plasma propagate outwards
from overdense regions until they are frozen out at recombination, when the 
baryons and photons decouple. These small enhancements in the density field
can be seen in the distribution of galaxies at later times. A BAO feature
can be seen in measurements of the two-point statistics of galaxies, at 
a characteristic comoving length scale of $\sim 100~\hMpc$, which can be used as a
standard ruler for measuring distances throughout cosmic time.

Redshift Space Distortions (RSD) can be used to measure the growth of structure.
Since we measure the redshift for each galaxy, the apparent position along the line of
sight depends on its peculiar velocity. The infall of galaxies towards overdense regions,
and the motion of galaxies within virialized haloes, leads to 
anisotropic distortions in the spatial distribution of galaxies in redshift space \citep{Kaiser1987}. 
RSD measurements can be used to measure the linear growth rate, \highlight{$f$, which in}
the \LCDM\ model, is related to the matter content of the Universe through 
$f \simeq \Omega_\mathrm{m}^\gamma$. In general relativity, $\gamma \simeq 0.55$, and
therefore, measurements of the growth rate allow constraints to be placed on theories
of modified gravity \citep[e.g.][]{Guzzo2008}.

The extended Baryon Oscillation Spectroscopic Survey \citep[eBOSS;][]{Dawson2016}, is
a survey of luminous red galaxies (LRGs), emission line galaxies (ELGs) and 
quasars, which is an extension of the previous BOSS survey \citep{Dawson2013}. 
\highlight{Between $0.8<z<2.2$, quasars are ideal tracers of the matter density field, since they are the brightest objects, 
and can be observed more easily than other types of galaxies}. This was first done in \citet{Ata2018}, filling in this redshift
range.

Two-point statistics, which are commonly used in cosmology, provide a way to quantify the distribution
of galaxies in a given volume. They measure the excess probability of finding pairs of galaxies,
as a function of spatial separation, compared to a random distribution \citep[e.g.][]{Landy1993}.
Measurements of the two-point statistics of galaxies in the survey can be used to
probe the growth rate and expansion history \citep[e.g.][]{Blake2011,Beutler2012,Alam2017}.
However, calculating cosmological distances from measurements of redshift requires a fiducial cosmology 
to be chosen. If this choice is incorrect, it will produce incorrect distance measurements,
resulting in a distortion of the BAO peak, which is known as the Alcock-Paczynski effect \citep{Alcock1979}. 
\highlight{By measuring the distortion of the BAO feature, which is quantified by the parameters
$\apar$ and $\aperp$ (parallel and perpendicular to the line of sight, respectively), 
constraints can be placed on the transverse comoving distance, $\dM(z)/\rdrag$, and the Hubble distance 
$\dH(z)/\rdrag$, where $\rdrag$ is the sound horizon at the drag epoch.}

In order to extract cosmological information from the survey, we need to be able to model the two-point statistics of QSOs. 
Our models, which are based on using perturbation theory to calculate cosmological power spectra or correlation functions \citep[e.g.][]{Crocce2006,Taruya2012,Carlson2013} 
are for dark matter, and neglect the formation
and evolution mechanisms of quasars. On large scales, this complicated non-linear, baryonic and strong gravity
physics is expected to decouple, and be modelled using a few nuisance parameters \citep{Hou2018,Zarrouk2018}.
It is therefore crucial to test the limits of such models, with realistic simulated quasar catalogues.

To validate the models used, we run our analysis on a set of N-body mock catalogues
in which the true cosmology is 
known. Measuring the scatter in the best fit parameters obtained with mocks
produced using different Halo Occupation Distributions (HODs), cosmologies, 
and including the effects
of redshift smearing and catastrophic redshifts, allows us to estimate 
the systematic uncertainties in these measurements.
Previously, quasar mock catalogues for the first year of eBOSS analysis were created in 
\citet{Rodriguez-Torres2017}, and multi-tracer mocks were created in \citet{Alam2019}.
The aim of this mock challenge is to firstly validate the RSD models on a set
of non-blind N-body mocks, with a wide variety of QSO models, 
where the true cosmology is known, and also on a set
of mocks where the cosmology is blinded. Secondly, we use these mocks to measure
the systematic uncertainties on $\fsig$, $\apar$ and $\aperp$.

\highlight{
The mock challenge presented in this paper is similar to what was previously done for the
BOSS survey, where N-body mocks, generated using a range of HOD models and also in different cosmologies, 
were analysed blindly \citep{Beutler2017}. 
For the WiggleZ survey \citep{Drinkwater2010}, the RSD analysis was performed using a range 
of existing models without explicitly testing them on mocks \citep{Blake2011RSD}.
In the cosmological analysis of the Dark Energy Survey Year 1 data \citep{Abbott2018}, the 
parameter measurements were validated using N-body mocks which simulated galaxy clustering
and lensing observables \citep{MacCrann2018}.
For future surveys, such as DESI \citep{DESI2016science,DESI2016instrument}, LSST \citep{Ivezic2019} 
and Euclid \citep{Laureijs2011}, much effort will be required in
producing many accurate mock catalogues to ensure that the model systematics can be reduced to the
required levels. 
}

The quasar mock challenge is part of the final release of BAO and RSD measurements from eBOSS.
The construction of the data catalogues is described in \citet{Ross2020,Lyke2020}, and
the configuration-space and Fourier-space analysis of the quasar sample is presented in \citet{Hou2020}
and \citet{Neveux2020}, respectively. In addition, cosmological measurements are made using the sample
of luminous red galaxies \citep[LRGs;][]{LRG_corr2020,Gil-Marin2020} and emission
line galaxies \citep[ELGs;][]{Raichoor2017,Tamone2020,DeMattia2020}, with mock catalogues and
mock challenges described in \citet{Alam2020,Avila2020,Lin2020,Rossi2020,Zhao2020}. At high redshifts, the BAO analysis of
Ly$\alpha$ forest measurements is found in \citet{duMasdesBourboux2020}.
The final cosmological implications of these results is presented in \citet{eBOSS_Cosmology2020}.\footnote{A summary of all SDSS BAO and RSD measurements with accompanying legacy figures can be found at \url{https://sdss.org/science/final-bao-and-rsd-measurements/}. The full cosmological interpretation of these measurements can be found at \url{https://sdss.org/science/cosmology-results-from-eboss/}}

This paper is outlined as follows. In Section~\ref{sec:eboss_qsos}, we give an overview
of the eBOSS QSO clustering sample, and observational effects.
In Section~\ref{sec:rsd_models}, we describe the RSD models used in the clustering analysis.
Section~\ref{sec:mock_catalogues} describes the HOD models, and methodology for creating
\highlight{the mocks.}
Results are discussed in Section~\ref{sec:non_blind_challenge} for the non-blind mocks,
and Section~\ref{sec:blind_challenge} for the blinded mocks. Our conclusions are summarised
in Section~\ref{sec:conclusions}.


\section{Quasar sample} \label{sec:eboss_qsos}

The eBOSS survey was performed as part of SDSS-IV \citep{Blanton2017}, the fourth phase of the Sloan Digital Sky Survey
\citep[SDSS;][]{York2000}. The survey, which began in July 2014 and completed observations in March 2019,
was carried out using the 2.5-metre Sloan Foundation Telecope at Apache Point Observatory in 
New Mexico \citep{Gunn2006}. The survey is the successor of the previous BOSS survey \citep[][which was 
part of SDSS-III]{Dawson2013}
and utilizes the two BOSS spectrographs \citep{Smee2013}, which are able to measure a total of $1000$ spectra per observation.

In order to make cosmological measurements covering a wide range of redshifts, the eBOSS survey targeted several
different types of tracers. These tracers are luminous red galaxies \citep[LRGs;][]{Prakash2016}, covering the redshift range $0.6<z<1.0$, 
emission line galaxies \citep[ELGs;][]{Raichoor2017}, over the redshift range $0.6<z<1.1$ and quasars \citep{Myers2015}. 
The quasars can be split into a lower redshift sample ($0.8<z<2.2$), which can be used as direct tracers of 
the matter field, and a higher redshift sample, which can be utilised for measurements of the 
Ly$\alpha$ forest ($z>2$). In this work we focus on the sample of quasars in the redshift range $0.8<z<2.2$
which can be used as direct tracers.

In the total sample, there are $\sim 330,000$ quasars, with $\sim 60\%$ in the northern galactic cap, and
$\sim 40\%$ in the south. The effective redshift of the quasar sample is $\zeff=1.48$,
covering a total area of $4700~\sqdeg$. See \citet{Ross2020} for details of the large-scale
structure catalogue, and \citet{Lyke2020} for details of the quasar catalogue, including the
procedure for determining redshifts. The number of quasars in the DR16 sample is approximately
doubled from the earlier DR14 sample. For the previous DR14 QSO analyses, see 
\citet{Gil-Marin2018, Hou2018, Zarrouk2018}.

Cosmological information can be extracted from the two-point clustering of the quasar redshift
catalogue. However, it is difficult to measure a precise redshift for quasars. 
There is some uncertainty in the redshift measurements, due to
astrophysical outflows. In addition, a small fraction of quasars have `catastrophic' redshift
measurements, where the redshift estimate is very far from the true redshift. 
It is essential that our models can deal with the impact this has on the clustering measurements
in order to obtain unbiased cosmological measurements.
We outline these effects below.

\subsection{Redshift smearing} \label{sec:redshift_smearing}

There is a wide range of behaviour in the width of the emission lines in the spectra of quasars.
Often, the optically selected quasars show broad emission lines, which is due to the fast rotation 
of hot gas close 
to the central black hole. The gas is also affected by radiation-driven winds, which leads to
offsets in the position of the emission lines in the spectra. This results in systematic
uncertainties in the measured quasar redshifts.

Several methods are used to estimate the redshift of quasars in the eBOSS survey. 
The pipeline redshifts are the result of fitting four eigenspectra to each spectrum.
The redshift can also be estimated from the MgII line, from a principle component 
analysis of the full quasar spectrum, or from visual inspection. The uncertainties
in the measured redshifts, i.e. redshift smearing, can be measured using the distribution
of $\Delta z$, where $\Delta z$ is the difference between two of these different
redshift estimates \citep[for details of the redshift estimates, see][]{Ross2020, Lyke2020}.

The quantity $\Delta z$ can equivalently be thought of as a velocity difference, $\Delta v$, 
where $\Delta v = c \Delta z / (1+z)$. The distribution of $\Delta v$ is approximately
a Gaussian, but with wide tails extending to large velocities \citep[see e.g. figure~4 of][]{Zarrouk2018}. 
In the survey requirements document, the
distribution of $\Delta v$ is a Gaussian with mean $\langle \Delta v \rangle = 0$,
and rms given by
\begin{equation}
\begin{aligned}
\sigma_v(z) &= 300~\kms               & z < 1.5 \\
\sigma_v(z) &= 450 (z-1.5) + 300~\kms & z > 1.5.
\end{aligned}
\label{eq:gaussian_smearing}
\end{equation}
The requirement of $300~\kms$ is relaxed for $z>1.5$, since it is more difficult to
obtain accurate redshifts. \highlight{We use this distribution as an initial model for the redshift
uncertainties that are applied in the mocks (see Section~\ref{sec:non_blind_mocks}), 
and then improve the modelling of redshift
smearing by looking at the distributions seen in the data.}

While the distribution of $\Delta v$ is approximately Gaussian, a Gaussian distribution is
unable to model the wide tails which extend to high velocities.
We investigated the distribution of $\Delta v$ by looking at the set of quasars which have duplicate
observations, and found that the distribution can be better modelled by the sum of two
Gaussians \citep[see figure~4 of][]{Lyke2020}.
This double-Gaussian distribution can be written as
\begin{equation}
\frac{dN}{d(\Delta v)} = \frac{1}{\sqrt{2\pi}(1+F)} \left[\frac{F}{\sigma_1} \exp\left( \frac{-\Delta v^2}{2 \sigma_1^2} \right) + \frac{1}{\sigma_2} \exp\left( \frac{-\Delta v^2}{2 \sigma_2^2} \right) \right],
\label{eq:realistic_smearing}
\end{equation}
where both Gaussians are centred on zero, with rms $\sigma_1$ and $\sigma_2$, and $F$ sets the
fraction of the two Gaussians. 
For objects which have been re-observed, the distribution of $\Delta v$ is fit well by
a double Gaussian probability distribution with $\sigma_1=150~\kms$, $\sigma_2=1000~\kms$, and $F=4.478$
\citep{Lyke2020}.

\subsection{Catastrophic redshifts} \label{sec:catastrophic_redshifts}

Measurements of quasar clustering are also affected by catastrophic redshifts, where an 
incorrect redshift is measured for a small fraction of objects, 
e.g. due to line confusion or contamination from sky lines. The catastrophic redshift
failure rate is low; for DR16 this is about 1.5\%. This is estimated from a set of 10,000
spectra which are randomly chosen for visual inspection. A castrophic redshift is defined
as having a pipline redshift which differs from the visual inspection redshift
by $\Delta v > 3000~\kms$. For more details, see \citet{Lyke2020}. 
The inclusion of catastrophic redshifts will impact the measurements of $\fsig$, since
the clustering signal is being diluted.

\subsection{Effective redshift}

\highlight{
There is also some uncertainty on the effective redshift of the QSO sample, as there is some
ambiguity in the definition of $\zeff$ \citep[see appendix~A of][]{Hou2020}. Changing the definition
shifts the effective redshift of the QSOs from $\zeff=1.48$ to $\zeff=1.52$. As shown in
\citet{Hou2020}, the impact this has on the cosmological measurements is small compared to the 
statistical error.
The effect of redshift smearing and catastrophic redshifts of $\zeff$ is negligible.
}


\section{Models for two-point statistics} \label{sec:rsd_models}

\highlight{
In this section we describe the RSD and BAO models that we test in this mock challenge.
Section~\ref{sec:two_point_statistics} gives an overview of two-point statistics in redshift space, 
and Section~\ref{sec:ap_effect} describes the Alcock-Paczynski effect.
The full-shape RSD models in Fourier space that we consider are described in 
Section~\ref{sec:fourier_models}, with configuration-space models in Section~\ref{sec:config_models}.
Finally, the BAO-only models, in both Fourier and configuration space, are described in
Section~\ref{sec:bao_models}.
}

\subsection{Two-point statistics}
\label{sec:two_point_statistics}

The galaxies and quasars observed in large cosmological surveys are biased
tracers of the underlying matter density field. The density contrast of tracers,
$\delta_t(\vect{x})$, at position $\vect{x}$ is related to the matter
overdensity, $\delta_m(\vect{x})$, by the linear bias, \highlight{$b_1$},
\begin{equation}
\delta_t(\vect{x}) = b_1 \delta_m(\vect{x}).
\label{eq:linear_bias}
\end{equation}
The two-point correlation function, $\xi(\vect{r})$ is defined as
\begin{equation}
\xi(\vect{r}) = \langle \delta(\vect{x}) \delta(\vect{x}+\vect{r}) \rangle,
\end{equation}
where the angled brackets indicate an ensemble average, and its Fourier
transform is the power spectrum, $P(\vect{k})$.

Galaxies and quasars are observed in redshift space \citep{Sargent1977}, and their apparent 
position along the line of sight is shifted by the peculiar velocity,
\begin{equation}
\vect{s} = \vect{x} + u_z \vect{\hat{z}}
\end{equation}
where $u_z$ is the component of velocity along the line of sight of the observer, 
expressed in units of the Hubble velocity, $u_z = v_z/aH$. The effect of velocity
is that it imparts anisotropies on the redshift-space two point correlation function, due to
two effects. The coherent infall of galaxies towards large overdensities leads to a flattening
in the clustering measurements on large scales, which is the Kaiser effect \citep{Kaiser1987}, while the motion
of galaxies within virialized haloes leads to Finger-of-God distortions \citep{Jackson1972}. 
In redshift space, the 2D correlation function, $\xi(s,\mu)$ can be decomposed into Legendre
multipoles,
\begin{equation}
\xi(s,\mu) = \sum_\ell \xi_\ell(s) \mathcal{L}_\ell(\mu),
\end{equation}
where $\mathcal{L}_\ell(\mu)$ is the $\ell^\mathrm{th}$ order Legendre polynomial, and $\mu$ is 
the cosine of the angle between the line of sight and the pair separation vector.
The multipoles $\xi_\ell(s)$ are
evaluated through
\begin{equation}
\xi_\ell(s) = \frac{2\ell + 1}{2} \int_{-1}^{1} \xi(s,\mu)\mathcal{L}_\ell(\mu) d\mu.
\label{eq:xi_multipoles}
\end{equation}
In linear theory, only the monopole, $\xi_0(s)$, quadrupole, $\xi_2(s)$, and hexadecapole, $\xi_4(s)$,
are non-zero.

\subsection{Alcock-Paczynski effect}
\label{sec:ap_effect}

\highlight{
On large scales, a BAO peak can be seen in the correlation function of galaxies (or alternatively oscillatory
features in the power spectrum), which is shifted if an incorrect fiducial cosmology
is assumed \citep{Alcock1979}. The BAO peak position is scaled by $s_\parallel' = \apar s_\parallel$ 
and $s_\bot' = \aperp s_\bot$
parallel and perpendicular to the line of sight, respectively, where
\begin{align}
\apar = \frac{H^\mathrm{fid}(z)\rdrag^\mathrm{fid}}{H(z) \rdrag}  && \mathrm{and} && \aperp = \frac{\dM(z)\rdrag^\mathrm{fid}}{\dM^\mathrm{fid}(z) \rdrag}.
\end{align}
$H(z)$ is the Hubble parameter, $\dM(z)$ is the transverse comoving distance, and 
$\rdrag$ is the sound horizon at the drag epoch. Quantities labelled `fid' are in the 
fiducial cosmology, while the true quantities which we aim to measure have no label.
Therefore, constraining the $\apar$ and $\aperp$ parameters from the two-point clustering
measurements can be used to place constraints on $\dM(z)/\rdrag$ and $\dH(z)/\rdrag$, 
where the Hubble distance $\dH(z) = c/H(z)$.
}

\subsection{Fourier space RSD models}
\label{sec:fourier_models}

In the linear approximation, the power spectrum in redshift space can be described using the
well-known Kaiser formula,
\begin{equation}
P_t^\mathrm{lin}(k,\mu) = (b_1^2 + 2fb_1\mu^2 + f^2\mu^4) P_m(k),
\label{eq:Pk_redshift_space_kaiser}
\end{equation}
\highlight{where $f$ is the linear growth rate, which is defined as
\begin{equation}
f(a)=\frac{d \ln D(a)}{d \ln a}, 
\label{eq:lin_growth_rate}
\end{equation}
$D$ is the linear growth function, and $a$ is the expansion factor \citep{Kaiser1987}. }
\highlight{This is only valid in the linear regime, where the density perturbations
are small ($\delta_m \ll 1$), and non-linear terms in the fluid equations describing the density
and velocity of the matter field are negligible. This corresponds to large physical scales 
($\gtrsim 50~\hMpc$), where the density of the galaxies
is related to matter density by a constant linear bias (Eq.~\ref{eq:linear_bias}), 
and there is a linear coupling between the matter velocity and density fields 
($\theta_m = f \delta_m$). Haloes and galaxies form in high density regions, 
on physical scales much smaller than $50~\hMpc$, where the, non-linear terms in the 
fluid equations are important, and the assumptions of linear theory break down. 
To predict the two-point statistics on small scales, it is therefore important to 
model RSD in the non-linear regime.}

The non-linear matter power spectrum for biased tracers, $P_t(k,\mu)$, is given by 
\begin{equation}
\begin{split}
P_t(k,\mu) &= P_{\delta \delta,t}(k) + 2f\mu^2 P_{\delta\theta,t}(k) + f^2\mu^4 P_{\theta \theta}(k) \\
&+ b_1^3 A(k,\mu) + b_1^4 B(k,\mu),
\end{split}
\label{eq:Pk_redshift_space}
\end{equation}
where $P_{\delta \delta, t}(k)$, $P_{\delta \theta,t}(k)$ and $P_{\theta \theta}(k)$ are the density,
density-velocity and velocity-velocity power spectra, respectively, for tracers, and 
$\theta=\vect{\nabla} \cdot \vect{u}$ is the divergence of the velocity field \citep{TNS2010}.
The first three terms in this expression are the non-linear Kaiser formula, 
$b_1$ is the linear bias, and $A(k,\mu)$ and $B(k,\mu)$ are 
correction terms. These correction terms arise from the non-linear coupling between the 
velocity and density fields, and depends on the cross-bispectrum, $P_{\delta \theta, t}(k)$ 
and $P_{\theta \theta}(k)$. The power spectrum and correction terms in Eq.~\ref{eq:Pk_redshift_space}
can be calculated using perturbation theory models.

On even smaller scales \highlight{($\lesssim 10~\hMpc$)}, the motion of galaxies within haloes 
becomes important, giving rise to Finger-of-God distortions. This can be modelled as
\begin{equation}
P_t^s(k,\mu) = F_\mathrm{FoG}(k,\mu) P_t(k,\mu) \exp\left[ - (k \mu \sigma_\mathrm{zerr})^2 \right].
\label{eq:pk_redshift_space_FoG}
\end{equation}
The first term, $F_\mathrm{FoG}(k,\mu)$, is a damping function which models the Finger-of-God effect,
and usually takes the form of a Gaussian or Lorentzian function. We use a function of the form
\begin{equation}
F_\mathrm{FoG}(k,\mu) = \frac{1}{\sqrt{1 + \mu^2 k^2 a_\mathrm{vir}^2}} \exp \left( \frac{-\mu^2 k^2 \sigma_v^2}{1 + \mu^2 k^2 a_\mathrm{vir}^2} \right),
\label{eq:FoG_damping}
\end{equation}
where $a_\mathrm{vir}$ is the kurtosis of the velocity distribution, and $\sigma_v$ is 
the velocity dispersion \citep{Grieb2017,Sanchez2017b,Hou2018}. 

The final term in Eq.~\ref{eq:pk_redshift_space_FoG} is an exponential function that models
the effect of redshift uncertainty, where $\sigma_\mathrm{zerr}$ is the redshift error \citep{Hou2018}.
\highlight{In the case of the double-Gaussian redshift smearing of Eq.~\ref{eq:realistic_smearing},
the parameters $\sigma_v$, $a_\mathrm{vir}$ and $\sigma_\mathrm{zerr}$ are able to pick out
the different Gaussian components (see Section~\ref{sec:model_stability}).}

In the next subsections, we briefly outline the specific Fourier space models we use.

\begin{table} 
\caption{\highlight{Parameters of the \Neveux\ (Section~\ref{sec:RegPT}), \Hou\ 
(Section~\ref{sec:respresso}) and CLPT (Section~\ref{sec:CLPT}) models. Prior ranges are
indicated for the free parameters.}}
\begin{tabular}{cccccc}
\hline
\multicolumn{2}{|c|}{\Neveux} & \multicolumn{2}{|c|}{\Hou} & \multicolumn{2}{|c|}{CLPT}\\
\hline
$b_1$ & $[0,5]$ &      $b_1$ & $[0.25,6]$ & $b$ & $[0,3]$\\
$b_2$ & $[-8,8]$ &      $b_2$ & $[-2,3]$ & $F''$& $[0,10]$\\
$b_\mathrm{s2}$ & fixed &   $\gamma_2$& fixed & & \\
$b_\mathrm{3nl}$ & fixed &  $\gamma_3^-$ & $[-2,2]$ & & \\
\hline
$A_g$ & $[-1,5]$ & & & & \\
\hline
$\sigma_v$ & $[0,15]$ & $\sigma_v$ & fixed & $\sigma$ & $[0,20]$ \\
$a_\mathrm{vir}$ & $[0,15]$ &  $a_\mathrm{vir}$ & $[0.2,10]$ & & \\
&  & $\sigma_\mathrm{zerr}$ & $[0,6]$ & & \\
\hline
\end{tabular}
\label{tab:model_parameters}
\end{table}

\subsubsection{RegPT} \label{sec:RegPT}

\highlight{In the first model that we consider,} the power spectra $P_{\delta\delta}(k)$,
$P_{\delta\theta}(k)$ and $P_{\theta\theta}(k)$, 
and the correction terms $A(k,\mu)$ and $B(k,\mu)$ of Eq.~\ref{eq:Pk_redshift_space}, 
\highlight{are calculated using} Regularized Perturbation Theory (RegPT), at 2-loop order \citep{Taruya2012}. 

The effect of redshift errors is treated as a velocity dispersion. The parameter $\sigma_v$ in the
Finger-of-God term is a free parameter, to take into account the effect of redshift errors, while the 
additional parameter $\sigma_\mathrm{zerr}$ is set to zero. 
\highlight{This choice implies that there is no additional damping
due to redshift errors.}

In the bias expansion, $b_1$ and $b_2$ are the linear and second order bias, 
and the local biases $b_{s2}$ and $b_{3\mathrm{nl}}$ are kept fixed assuming local Lagrangian
bias. There is also an additional shot noise parameter, $A_g$ \citep{Neveux2020}.
\highlight{In addition to $\apar$, $\aperp$ and $\fsig$, there are 5 free parameters
($b_1$, $b_2$, $A_g$, $a_\mathrm{vir}$, and $\sigma_v$), and their priors are given in Table~\ref{tab:model_parameters}.}

\highlight{The model we have described here has previously been used in the analysis of 
BOSS LRGs \citep{Beutler2014, Beutler2017}. At $z=1$, the power spectrum prediction using RegPT is 
within percent-level agreement of N-body simulations up to $k \sim 0.23~\invhMpc$ \citep{Taruya2012}.}

\highlight{
In the final eBOSS DR16 quasar analysis, the same model is fit to measurements of the power spectrum multipoles 
in Fourier space \citep{Neveux2020}, and they are therefore able to use the systematic uncertainties quoted 
in this study. This model is additionally used in the Fourier space analysis of the eBOSS ELGs \citep{DeMattia2020}. }
Throughout this paper, we will refer to
this model, which combines RegPT with the specific FoG prescription described above, as simply `RegPT'.

\subsubsection{RESPRESSO + Fitting Formula}
\label{sec:respresso}

\highlight{In the second model we consider,} 
the power spectrum $P_{\delta \delta,t}(k)$ is computed using the code \textsc{respresso} \citep{Nishimichi2017}.
\textsc{respresso} is based on the response function, which characterises how the non-linear power spectrum
varies in response to small perturbations of the initial power spectrum.
$P_{\delta \theta,t}(k)$ and $P_{\theta \theta,t}(k)$ are calculated 
from the fitting functions of \citet{Bel2019}, which are based on measurements from N-body simulations.

In the Finger-of-God term of Eq.~\ref{eq:FoG_damping}, $\sigma_v$ is kept fixed to the linear
theory prediction. However, $\sigma_v$ and $a_\mathrm{vir}$ are defined differently than in the
\Neveux\ model described in Section~\ref{sec:RegPT}, differing by a factor of $f$ in their
normalization. The effect of redshift errors is modelled by the parameter $\sigma_\mathrm{zerr}$,
which is kept free.

The bias expansion of \citet{Chan2012} is used \citep[see equation~22 of][]{Hou2020},
with first and second order bias parameters $b_1$ and $b_2$, and non-local bias parameters
$\gamma_2$ and $\gamma^-_3$. $\gamma_2$ is fixed assuming local Lagrangian bias. 
In total, there are 5 free parameters  
($b_1$, $b_2$, $\gamma^-_3$, $a_\mathrm{vir}$ and $\sigma_\mathrm{zerr}$), with
priors in Table~\ref{tab:model_parameters}.

\highlight{While this model gives a prediction of the redshift-space power spectrum, the two-point correlation
function can by calculated from its Fourier transform. A similar model to this has previously been used in 
the DR14 QSO analysis in configuration space of \citet{Hou2018} 
\citep[and also in the analysis of the BOSS wedges][]{Grieb2017,Sanchez2017b}
but this differs in that the power spectra were calculated using Galilean-invariant Renormalized 
Perturbation Theory (gRPT). The \textsc{respresso} prediction of the matter power spectrum is
in good agreement with N-body simulations, within 2\% up to $k=0.3~\invhMpc$, 
and shows a slight improvement compared to gRPT \citep[see figure~4 of][]{Hou2020}. 
In addition, \citet{Hou2018} showed that cosmological parameter estimates are less biased with
the parameter $z_\mathrm{err}$ included.
}

\highlight{The full model as presented here, which we measure the systematic errors of,
is used in the configuration space DR16 quasar analysis of \citet{Hou2020}. We will refer
to this model as `\Hou'.}

\subsection{Configuration space RSD models}
\label{sec:config_models}

An alternative approach is to model the two-point correlation function in configuration
space.

\subsubsection{CLPT} \label{sec:CLPT}

In the Lagrangian approach, particles (or tracers) are moved from their initial Lagrangian 
coordinates, $\vect{q}$, to their final coordinates $\vect{x}$, by
a displacement field, $\vect{\psi}$,
\begin{equation}
\vect{x}(\vect{q},t) = \vect{q} + \vect{\psi}(\vect{q},t)
\label{eq:displacement_field}
\end{equation}

The displacement field can be written as a perturbative expansion, 
$\vect{\psi}=\vect{\psi}^{(1)} + \vect{\psi}^{(2)} + \vect{\psi}^{(3)} + \cdots$, 
where the first order term is the Zel'dovich approximation. 
In the Convolution Lagrangian Perturbation Theory (CLPT) model,
more terms are summed together in this expansion, which provides a good
description for the real space correlation function \citep{Matsubara2008,Carlson2013}.

The real space correlation function can then be transformed into redshift space using 
the Gaussian streaming model,
\begin{equation}
1+\xi(s_\bot,s_\parallel) = \int dr_\parallel (1 + \xi(r)) \mathcal{G}(s_\parallel-r_\parallel, v_{12}, \sigma_{12})
\end{equation}
where $\mathcal{G}$ is a Gaussian function centred on $\mu v_{12}$, which describes the 
probability that a pair of galaxies with separation $r_\parallel$ in real space
have a separation $s_\parallel$ in redshift space,
\begin{equation}
\mathcal{G}(s_\parallel-r_\parallel, v_{12}, \sigma_{12}) = \frac{1}{\sqrt{2 \pi \sigma_{12}^2(r,\mu)}} \exp \left( \frac{(s_\parallel - r_\parallel - \mu v_{12})^2}{2\sigma_{12}^2(r,\mu)} \right).
\end{equation}
This probability depends on the velocities of galaxies; $v_{12}$ is the pairwise infall velocity,
and $\sigma_{12}$ is the pairwise velocity dispersion. 

In total there are 3 nuisance parameters, which are
the linear and second order bias, $b$ and $F''$ respectively, and $\sigma$, which takes into account the 
Finger-of-God effect and redshift smearing. \highlight{Their priors are given in Table~\ref{tab:model_parameters}.}

\highlight{
The CLPT prediction for the redshift-space correlation function is fit to measurements of the
correlation function in real space. Previously, this model was used in the eBOSS
DR14 quasar analysis of \citet{Zarrouk2018} in configuration space, and also in 
\citet{Icaza-Lizaola2019} for the LRGs. CLPT was also used in the BOSS LRG analysis of \citet{Satpathy2017}.
At the redshift of the QSOs, the model provides accurate correlation function predictions which 
agree with N-body simulations down to small scales of $\sim 20~\hMpc$ \citep{Zarrouk2018}.
}

\highlight{The CLPT model is used in the eBOSS DR16 LRG analysis of \citet{LRG_corr2020}, but is not
used in the final analysis of the QSOs.}
For the analysis of the DR16 quasar sample, the \Neveux\ model (Section~\ref{sec:RegPT}) and
the \Hou\ model (Section~\ref{sec:respresso}) are used in Fourier and configuration
space, respectively. As we show in this paper,
the three models work equally well, and there is no reason why the CLPT model should be chosen to
be used in the final analysis over any of the other models that we consider. 
In this work we use the CLPT model as an additional check to verify the rescaling of the blinded 
mocks (Section~\ref{sec:blind_mocks}) independently of the other models.

\subsection{BAO models}
\label{sec:bao_models}

In addition to the RSD models described in the sections above, as part of the mock
challenge we also verify the BAO fitting procedure, which provides another way
to measure $\apar$ and $\aperp$ from the two-point statistics of the data. 
BAO modelling aims to isolate cosmological information from the BAO only, which differs from 
RSD modelling, which predicts the two-point statistics over a wide range of scales.
In \citet{Neveux2020}, BAO fits are performed in Fourier space on measurements
of the power spectrum, while in \citet{Hou2020}, the fits are done to the correlation
function measurements in configuration space.
We describe the BAO models below.

\subsubsection{Fourier space analysis}

The linear power spectrum can be decomposed into two parts \citep[e.g.][]{Kirkby2013}: a `smooth' (or `no-wiggles') 
broadband term, $P_\mathrm{sm}(k)$, and `peak' term, $P_\mathrm{peak}(k)$, 
which isolates the oscillations of the BAO feature,
\begin{equation}
P_\mathrm{lin}(k) = P_\mathrm{sm}(k) + P_\mathrm{peak}(k).
\end{equation}
\highlight{In the first BAO model we consider,} the redshift-space power spectrum, 
which models non-linear effects, is given by
\begin{equation}
P(k,\mu) =  \frac{b^2(1+\beta \mu^2)^2}{1 + (k\mu \Sigma_\mathrm{s})^2/2} \left[ P_\mathrm{sm}(k,\mu) + P_\mathrm{peak}(k,\mu) e^{-k^2\Sigma_\mathrm{nl}^2}\right],
\label{eq:Pk_BAO_model}
\end{equation}
where $b$ is the bias, and $\beta=f/b$ is the redshift space distortion parameter \citep{Bautista2018}. 
The factor of $\big(b^2(1+\beta \mu^2)^2\big) / \big(1 + (k\mu \Sigma_\mathrm{s})^2/2\big)$ 
is a Lorentzian which models the Finger-of-God effect and redshift smearing 
on small scales, with the damping parameter $\Sigma_\mathrm{s}$. The factor of 
$b^2(1+\beta \mu^2)^2$ comes from the Kaiser formula (Eq.~\ref{eq:Pk_redshift_space_kaiser}).
In Eq.~\ref{eq:Pk_BAO_model}, $P_\mathrm{peak}(k,\mu)$ is multiplied by an exponential
term, which models the anisotropic, non-linear damping of the BAO feature \citep{Eisenstein2007}.
This anisotropic damping parameter, $\Sigma_\mathrm{nl}$, is defined as
\begin{equation}
\Sigma_\mathrm{nl}^2 = (1-\mu^2)\Sigma_\bot^2/2 + \mu^2 \Sigma_\parallel^2/2.
\label{eq:sigma_nl}
\end{equation}

\highlight{The mock catalogues are used to obtain values of the damping parameters.}
For mocks which do not contain redshift smearing, the damping parameters are kept fixed in the fitting procedure
to the values $\Sigma_\mathrm{s}=1~\hMpc$, $\Sigma_\bot=3~\hMpc$ and $\Sigma_\parallel=8~\hMpc$.
These values are obtained from fits to the mocks with no redshift smearing (see Section~\ref{sec:non_blind_mocks}). 
For mocks which do contain redshift smearing, the value of $\Sigma_\mathrm{s}$ is
increased to $\Sigma_\mathrm{s}=4~\hMpc$.
This value of $\Sigma_s$ is obtained from mocks with realistic smearing, with the values of 
$\Sigma_\parallel$ and $\Sigma_\bot$ fixed.

The broadband component of the power spectrum multipoles, which doesn't contain BAO information, 
is fit by a polynomial, $P^f_\ell(k) = P_\ell(k) + a_{0,\ell}/k + a_{1,\ell} + a_{2,\ell}k$,
where $a_{\ell,i}$ are the polynomial coefficients \citep{Bautista2018}. 
In total there are 10 nuisance parameters in this BAO model
(the bias, and 3 broadband terms for each multipole).

\highlight{The BAO model described in this section has previously been used in the analysis
of the eBOSS DR14 LRGs \citet{Bautista2018}. In \citet{Neveux2020}, this model is 
fit in Fourier space to measurements of the power spectrum of the DR16 QSOs.}

\subsubsection{Configuration space analysis}

\highlight{In the second BAO model we use}, the power spectrum in redshift space is given by
\begin{equation}
P(k,\mu) = \left( \frac{1+\beta\mu^2}{1 + (k\mu\Sigma_s)^2/2} \right)^2 \left[ P_\mathrm{sm}(k,\mu) + P_\mathrm{peak}(k,\mu)e^{-k^2 \Sigma_\mathrm{nl}^2}\right],
\end{equation}
which differs from Eq.~\ref{eq:Pk_BAO_model} by the Lorentian function, which models the 
Fingers-of-God distortions, being squared \citep{Ross2017}. The anisotropic damping parameter, 
$\Sigma_\mathrm{nl}$ is defined the same way (Eq.~\ref{eq:sigma_nl}).

$P_\mathrm{lin}$ is calculated using CAMB \citep{Lewis1999}, while the smooth `no-wiggles' power spectrum
is calculated from the fitting formulae of \citet{Eisenstein1998}. The damping parameters $\Sigma_\mathrm{s}$, $\Sigma_\bot$ and $\Sigma_\parallel$. are kept fixed to the same
values as are used for the Fourier space BAO model.

The broadband parts of the correlation function monopole and quadrupole are each modelled as a 
cubic polynomial, with two additional parameters that adjust the BAO feature
\citep[for details, see section~5 of][]{Hou2020}. In total, there are 8 nuisance parameters.

\highlight{
The model $P(k,\mu)$ is Fourier transformed to obtain the correlation function.
BAO fits using this model in configuration space were previously done in the analysis of the 
BOSS LRGs \citep{Ross2017}. The same model is used in the DR16 QSO correlation function BAO fits
of \citet{Hou2020}.}


\begin{figure} 
\centering
\includegraphics[width=0.95\linewidth]{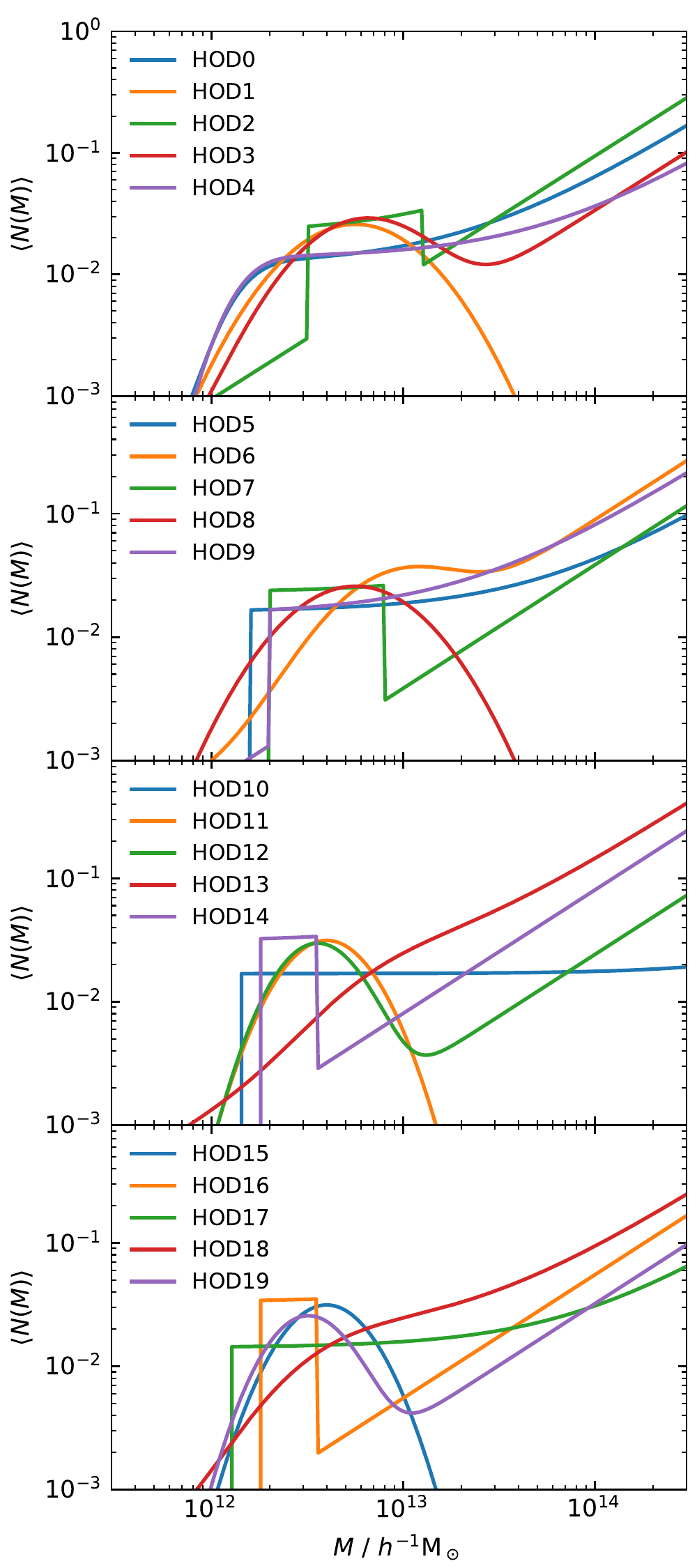}
\caption{The total halo occupation function for all 20 HOD models used to construct the non-blind 
quasar mock catalogues. For clarity, the models are split into 4 panels, where the HOD model
is indicated by the colour, given in the legend.}
\label{fig:hod}
\end{figure}

\begin{table*} 
\caption{HOD parameters used for each of the mock catalogues. The shape of the HOD for central quasars
is either a smooth step function (Sm), Gaussian (G), top hat function (TH) or a sharp step (Sh), while
satellites follow either a power law (PL) or gaussian (G) distribution. Satellites are also positioned
in the halo using dark matter particles from the simulation (Par), or following a NFW profile.
The parameter $f_\mathrm{sat}$ is the satellite fraction, and $\tau$ is the quasar duty cycle. 
$\log M_\mathrm{cen}$ is the central position of the HOD function for central quasars.
$\log \sigma_M$ sets the width of smooth step function, or Gaussian HODs, while $\Delta \log M$ 
is the width of the top hat HOD. $M_\mathrm{sat}$ is the low mass cutoff in the power law for satellites, 
and $\alpha_\mathrm{sat}$ is the power law slope. `Sat. cut' indicates whether the central HOD
was used to add the low mass cutoff in the HOD of satellites. The final column, $n$, is the number
density of quasars, in units $10^{-5} (\hMpc)^{-3}$.}
\begin{tabular}{cccccccccccccc}
\hline
HOD & Cen. & Sat. & Sat. & $f_\mathrm{sat}$ & $\tau$ & $\log M_\mathrm{cen}$ & $\log \sigma_M$ & $\Delta \log M$ & $M_\mathrm{sat}$ & $M_\mathrm{cut}$ & $\alpha_\mathrm{sat}$ & Sat. & $n/10^{-5} $ \\
    & HOD & HOD & pos. & & & & & & & & & cut & $(\hMpc)^{-3}$ \\
\hline
\textsc{hod0}  & Sm & PL & Par & 0.19  & 0.012 & 12.13 & 0.2 & -   & 15.29 & 11.61 & 1.0 & No  & 2.185 \\
\textsc{hod1}  & G  & G  & NFW & 0.07  & 0.074 & 12.75 & 2.0 & -   & -     & -     & -   & -   & 2.186 \\
\textsc{hod2}  & TH & PL & NFW & 0.60  & 0.022 & 12.80 & -   & 0.6 & 15.03 & 10.57 & 1.0 & No  & 2.183 \\
\textsc{hod3}  & G  & PL & Par & 0.21  & 0.020 & 12.80 & 0.3 & -   & 15.47 & 10.57 & 1.0 & No  & 2.183 \\
\textsc{hod4}  & Sm & PL & NFW & 0.08  & 0.014 & 12.13 & 0.2 & -   & 15.64 & 11.61 & 1.0 & No  & 2.195 \\
\textsc{hod5}  & Sh & PL & NFW & 0.17  & 0.016 & 12.20 & -   & -   & 15.57 & 10.57 & 1.0 & No  & 2.184 \\
\textsc{hod6}  & G  & PL & NFW & 0.56  & 0.021 & 13.00 & 0.3 & -   & 15.05 & 10.57 & 1.0 & No  & 2.201 \\
\textsc{hod7}  & TH & PL & Par & 0.24  & 0.023 & 12.60 & -   & 0.6 & 15.41 & 10.57 & 1.0 & No  & 2.202 \\
\textsc{hod8}  & G  & G  & NFW & 1.00  & 0.074 & 12.75 & 2.0 & -   & -     & -     & -   & -   & 2.186 \\
\textsc{hod9}  & Sh & PL & NFW & 0.42  & 0.015 & 12.30 & -   & -   & 15.18 & 10.57 & 1.0 & No  & 2.197 \\
\textsc{hod10} & Sh & PL & Par & 0.002 & 0.017 & 12.15 & -   & -   & 15.36 & 10.57 & 1.0 & Yes & 2.181 \\
\textsc{hod11} & G  & G  & NFW & 0.10  & 0.060 & 12.60 & 1.6 & -   & -     & -     & -   & -   & 2.184 \\
\textsc{hod12} & G  & PL & NFW & 0.05  & 0.014 & 12.55 & 0.2 & -   & 15.61 & 10.57 & 1.0 & Yes & 2.188 \\
\textsc{hod13} & Sm & PL & NFW & 0.73  & 0.016 & 12.80 & 0.4 & -   & 14.89 & 10.57 & 1.0 & No  & 2.183 \\
\textsc{hod14} & TH & PL & Par & 0.17  & 0.031 & 12.40 & -   & 0.3 & 15.09 & 10.57 & 1.0 & Yes & 2.174 \\
\textsc{hod15} & G  & G  & Par & 0.50  & 0.060 & 12.60 & 1.6 & -   & -     & -     & -   & -   & 2.185 \\
\textsc{hod16} & TH & PL & NFW & 0.12  & 0.033 & 12.40 & -   & 0.3 & 15.23 & 10.57 & 1.0 & Yes & 2.186 \\
\textsc{hod17} & Sh & PL & NFW & 0.04  & 0.014 & 12.10 & -   & -   & 13.93 & 10.57 & 1.0 & Yes & 2.198 \\
\textsc{hod18} & Sm & PL & NFW & 0.36  & 0.018 & 12.50 & 0.4 & -   & 15.12 & 10.57 & 1.0 & No  & 2.182 \\
\textsc{hod19} & G  & PL & Par & 0.07  & 0.012 & 12.50 & 0.2 & -   & 15.49 & 10.57 & 1.0 & Yes & 2.187 \\
\hline
\end{tabular}
\label{tab:hods}
\end{table*}

\section{Mock catalogues} \label{sec:mock_catalogues}

In this work, we construct the mock catalogues using the OuterRim N-body 
simulation \citep{Habib2016,Heitmann2019a,Heitmann2019b}, which contains 10,240$^3$ 
particles of mass 
$m_p=1.85\times 10^9~\hMsun$ in a box of side length $3000~\hMpc$. Haloes 
are identified using a friends-of-friends (FOF) algorithm \citep{Davis1985}
with linking length $b=0.168$. The OuterRim simulation uses a flat $\Lambda$CDM
cosmology with $\Omega_\mathrm{cdm} h^2=0.1109$,
$\Omega_\mathrm{b} h^2=0.02258$, $h=0.71$, $\sigma_8=0.8$ and $n_s=0.963$,
which is consistent with the WMAP7 measurements \citep{Komatsu2011}.
Mock catalogues are constructed from the simulation snapshot at $z=1.433$. This 
snapshot is chosen as it is closest in redshift to the effective
redshift of the eBOSS quasar sample ($\zeff=1.48$).
\highlight{Using a single snapshot allows the accuracy of the models to be tested
at a single redshift, without evolution. However, in the mocks we construct, we also add 
in evolution in the velocities, testing our models with $f$ that evolves with redshift.}

Quasar clustering measurements can be interpreted using the halo model, 
in which quasars reside within dark matter haloes. The clustering
on large scales is described by a two-halo term, from pairs of quasars that 
occupy different haloes, while the one-halo term on small scales is from pairs
residing within the same halo. This link between quasars and their host haloes can be modelled using
a halo occupation distribution (HOD). 
Recent work has aimed to measure the HOD and satellite fraction of quasars, from
both observations and simulations \citep[e.g.][]{Leauthaud2015,Powell2018,Georgakakis2019,Oogi2019}.
Precise measurements of the small-scale quasar clustering \citep{Mohammad2020}, which is dominated
by the 1-halo term, will help to further constrain the models.
In this mock challenge, in order to place a conservative upper bound on the systematic
error in the measurements from the RSD models, we create mock catalogues covering a wide variety 
of different HODs.

We also aim to test the models on a range of different cosmologies, since the true cosmology
is also not known. The OuterRim simulation is a single N-body simulation, which was run in a 
cosmology that is consistent with the WMAP7 cosmological parameters.
To generate mocks in different cosmologies, it is not feasible to run many 
N-body simulations with the volume and resolution of OuterRim. Instead the halo catalogue of the 
original OuterRim simulation can be modified in order to mimic a catalogue of a different cosmology.

In this section, we describe the HOD models used, and methodology for creating mocks
in different cosmologies.

\subsection{HOD Modelling} \label{sec:hods}

The HOD describes the average number
of central and satellite quasars residing in haloes as a function of the halo mass, $M$.
The total number of quasars per halo is given by the sum of central and satellite
quasars,
\begin{equation}
\langle N_\mathrm{tot}(M) \rangle = \langle N_\mathrm{cen}(M) \rangle + \langle N_\mathrm{sat}(M) \rangle.
\end{equation}
Since quasars are rare, the probability that more than one quasar resides within the same
dark matter halo is low.

We construct mock catalogues using 5 different functional forms for the HODs. 
This allows us to explore a wide range of HODs, and test the impact of the 
HOD on the model fits.

\subsubsection{Smooth step and power law}

The first HOD model we consider is the same as was used for the 
quasar mocks in \citet{Zarrouk2018}, and uses a 5+1 parameter HOD 
\citep[e.g.][]{Tinker2012}, which is motivated by a monotonic relation
between quasar luminosity and host halo mass, with the brightest quasars residing
in the most massive haloes. The probability
that a halo contains a central quasar is given by the smooth step function
\begin{equation}
\langle N_\mathrm{cen}(M) \rangle = \tau \frac{1}{2} \left[1 + \mathrm{erf} \left( \frac{\log M - \log M_\mathrm{cen}}{\log \sigma_\mathrm{m}} \right) \right],
\end{equation}
where the position of this step is set by $M_\mathrm{cen}$.
The probability that a halo with mass $M \ll M_\mathrm{cen}$ hosts a central quasar is zero,
which transitions to a probability of $\tau$ for $M \gg M_\mathrm{cen}$. 
$\tau$ is the quasar duty cycle, which takes into account that not all \highlight{central black holes} are active.
\highlight{This is defined as the fraction of haloes which host an active central galaxy, and sets the height
of the step function. In our HOD models, the duty cycle is assumed to be constant, while observations show
a dependence on redshift and stellar mass of the host galaxy \citep[e.g.][]{Bongiorno2016,Georgakakis2017}.}
The width of the transition is set by the parameter $\log \sigma_\mathrm{m}$. 
This softening of the step function accounts for scatter in the relation between quasar
luminosity and halo mass.

The number of satellite quasars in each halo is Poisson distributed, 
with mean given by a power law,
\begin{equation}
\langle N_\mathrm{sat}(M) \rangle = \left( \frac{M}{M_\mathrm{sat}} \right)^{\alpha_\mathrm{sat}} \exp \left( - \frac{M_\mathrm{cut}}{M} \right),
\label{eq:hod_satellite_power_law}
\end{equation}
where $\alpha_\mathrm{sat}$ is the slope of the power law, $M_\mathrm{sat}$ 
sets the normalisation, and $M_\mathrm{cut}$ is a cutoff at low masses.

\subsubsection{Gaussian}

This HOD model is the same as was used in \citet{Kayo2012,Eftekharzadeh2019},
in which the total HOD for all quasars is given by a Gaussian
\begin{equation}
\langle N_\mathrm{tot}(M) \rangle = \frac{\tau}{\sqrt{2 \pi} \log\sigma_\mathrm{m}} \exp\left[ - \frac{(\log M - \log M_\mathrm{cen})^2}{2(\log\sigma_\mathrm{m})^2} \right].
\label{eq:hod_gaussian}
\end{equation}
This model is motivated by haloes in a narrow mass range hosting a wide range of quasar luminosities. 
The parameter which sets the mean of the Gaussian, $\log M_\mathrm{cen}$,
is the halo mass which is most likely to host quasars, and $\log\sigma_\mathrm{m}$ 
sets the width of the Gaussian. As with the previous HOD model, $\tau$ is the quasar duty cycle. 
The parameters $M_\mathrm{m}$,
$f_\mathrm{N}$ and $\Delta_\mathrm{m}$ in equation~6 of \citet{Eftekharzadeh2019} are related to the
parameters in Eq.~\ref{eq:hod_gaussian}:
$M_\mathrm{m}$ = $M_\mathrm{cen}$, $f_\mathrm{N} =\sqrt{\ln(10)}\tau$ and 
$\Delta_\mathrm{m} = \ln(10) \log\sigma_\mathrm{m}$.

The satellite fraction, $f_\mathrm{sat}$, is an additional parameter used to split the total
occupation distribution into the central and satellite HODs. For central quasars, 
$\langle N_\mathrm{cen}(M) \rangle = (1-f_\mathrm{sat})\langle N_\mathrm{tot}(M) \rangle$, 
while for satellites,
$\langle N_\mathrm{sat}(M) \rangle = f_\mathrm{sat}\langle N_\mathrm{tot}(M) \rangle$.

\subsubsection{Gaussian and power law}

We also consider a HOD model where the occupation function for central
quasars is given by a Gaussian, but with a power law for 
satellites. The central HOD has the same form as Eq.~\ref{eq:hod_gaussian}, and
the satellite occupation function is given by Eq.~\ref{eq:hod_satellite_power_law}.
This model, and the remaining HOD models we consider, are not physically motivated, but are
included in order to test our models over a wide range of different HODs.

\subsubsection{Sharp step and power law}
In this HOD model, the central occupation function is given by a sharp step function,
\begin{equation}
\langle N_\mathrm{cen}(M) \rangle = \tau \Theta(M-M_\mathrm{cen}),
\end{equation}
where $\Theta(x)$ is the Heaviside step function, $M_\mathrm{cen}$ sets the position of the step,
and $\tau$ is the quasar duty cycle. For satellites, the occupation function is the same power
law function as Eq.~\ref{eq:hod_satellite_power_law}.

\subsubsection{Top hat and power law}
The final HOD uses a top hat function for the central quasars, given by
\begin{equation}
\langle N_\mathrm{cen}(M) \rangle = \tau \Theta(\log\sigma_\mathrm{m} - |M - M_\mathrm{cen}|),
\end{equation}
where $\Theta(x)$ is the Heaviside step function, $M_\mathrm{cen}$ sets the central position of
the top hat, $\log\sigma_\mathrm{m}$ is the width, and $\tau$ is the duty cycle. As with
the previous HODs, the satellite occupation function is given by the power law of 
Eq.~\ref{eq:hod_satellite_power_law}.

HOD parameters for the 20 HOD models we use are given in Table~\ref{tab:hods}.
When using the HODs to populate haloes, we assume that the probability that a 
halo contains a central or satellite quasar is independent, making it possible for a halo
to contain satellite quasars with no central. In the HODs described above, the power
law for satellites can extend to low masses, below any cutoff in the centrals. For
some of the HOD models we use, the power law is continued to low masses. For others, we
multiply the satellite HOD by the central HOD, so the satellites have the same 
cutoff as the centrals. 
The various HOD models described above are illustrated in Fig.~\ref{fig:hod}. For each of the 5
HOD models, 4 sets of HOD parameters are used, to give a total of 20 HODs. 

\highlight{We have considered a wide range of HOD models, some of which are not physically
motivated (the models with sharp cuts). While this could potentially inflate the final
uncertainties that we measure for the different RSD models, we find that there is no strong
dependence in our results on the particular HOD model (see Section~\ref{sec:non_blind_results}).}

\highlight{
The 20 chosen sets of HOD parameters are tuned to produce approximately 
the same large-scale clustering and quasar number density. This tuning of the HOD
parameters is done so that the clustering on scales $s>20~\hMpc$ is in agreement
with the DR16 clustering measurements, and the number density of quasars in all mocks
is approximately the same, which is close to $2\times 10^{-5} (\hMpc)^{-3}$.
}

\highlight{In the following sections, we describe the methodology for creating the 
non-blind and blind mocks for this mock challenge. Table~\ref{tab:mocks_summary}
summarizes all the mocks that have been produced in this work.}

\begin{table*} 
\caption{\highlight{Summary of all the mock catalogues constructed for the quasar mock
challenge. This includes the non-blind mocks, the blind mocks with fixed HOD, 
and the blind mocks where the HOD is varied. For each, $N_\mathrm{tot}$ is the total
number of sets of 100 mocks, $N_\mathrm{HOD}$ is the number of unique HODs used, and $N_\mathrm{cosmo}$
is the number of unique cosmologies. We also indicate whether observational effects 
(redshift smearing and catastrophic redshifts) are included, the location of tables where more information
about the individual mocks can be found, and the sections which describe the construction of the mocks 
and discuss the results.}}
\begin{tabular}{cccccccc}
\hline 
Mocks & $N_\mathrm{tot}$ & $N_\mathrm{HOD}$ & $N_\mathrm{cosmo}$ & Obs. effects & Table & Construction & Results \\
\hline
Non-blind         & 20 & 20 & 1 & Yes & Table~\ref{tab:hods} & Section~\ref{sec:non_blind_mocks_method} & Section~\ref{sec:non_blind_results} \\
Blind (Fixed HOD) & 8  & 1  & 8 & No & Table~\ref{tab:cosmologies} & Section~\ref{sec:blind_mocks_method} & Section~\ref{sec:blind_results_fixed_hod} \\
Blind             & 24 & 7  & 8 & No & Table~\ref{tab:blind_mocks} & Section~\ref{sec:blind_mocks_method} & Section~\ref{sec:blind_results_varying_hod} \\
\hline
\end{tabular}
\label{tab:mocks_summary}
\end{table*}

\subsection{Non-blind mocks}
\label{sec:non_blind_mocks}

\highlight{
The mock catalogues for the non-blind part of the mock challenge are constructed by 
populating the OuterRim simulation snapshot at z=1.433 with quasars using the 20 HOD models 
outlined in the previous section. The effects of redshift smearing and catastrophic redshifts
are then applied to the mocks. These observational effects are described in 
Section~\ref{sec:observational_effects}, and the full methodology for creating the mocks
is given in Section~\ref{sec:non_blind_mocks_method}.
}

\subsubsection{Observational effects} \label{sec:observational_effects}

To test that the models are robust against the effects of redshift smearing 
(Section~\ref{sec:redshift_smearing}) and catastrophic redshifts (Section~\ref{sec:catastrophic_redshifts}),
we consider 4 different cases of mocks.

The `no smearing' mocks do not contain the effects of redshift smearing or catastrophic redshifts. 

The `Gaussian smearing' mocks include the effect of redshift smearing, where the redshift of each quasar is
shifted by a random $\Delta z$, which corresponds to a velocity shift, $\Delta v$ 
(Section~\ref{sec:redshift_smearing}). $\Delta v$ is drawn from the redshift-dependent Gaussian distribution 
(Eq.~\ref{eq:gaussian_smearing}), which is defined in the survey requirements document. 

We also create mocks with `realistic smearing', where for each quasar, $\Delta v$ is randomly drawn from a 
double-Gaussian distribution (Eq.~\ref{eq:realistic_smearing}). This distribution includes the wide tails,
extending to high velocities, which is seen in the data.
We keep the shape of this distribution fixed with redshift. Including this double-Gaussian
redshift smearing in the mocks will test whether the models can recover the expected parameters 
with a realistic redshift smearing distribution. The mocks with Gaussian redshift smearing will test the
models for the effect of the degradation of redshift accuracy at high redshifts.

The `catastrophic redshifts' mocks are identical to the `realistic smearing' mocks, but in addition include the
effect of catastrophic redshifts. A random 1.5\% of objects 
(which matched the catastrophic redshift failure rate of the data) 
are assigned a new redshift, which is drawn from a uniform distribution in $z$, such that the object remains
inside the cubic box.

\subsubsection{Methodology for creating non-blind mocks}

\label{sec:non_blind_mocks_method}

To construct mocks from the OuterRim simulation, \highlight{the z=1.433 snapshot is populated with quasars
using our HOD models}. For each halo, with mass $M$, a uniform random number $0<x<1$ is drawn, 
and the halo is chosen to contain a central quasar if $x < \langle N_\mathrm{cen} (M) \rangle$. 
The number of satellites in each halo is
drawn from a Poisson distribution with mean given by $\langle N_\mathrm{sat} (M)\rangle$. The central
and satellite HODs are independent of each other, so it is possible for a halo to contain a satellite
quasar with no central.

Central quasars are positioned at the centre of each halo, and are assigned the same velocity. 
Satellite quasars are positioned around the centre of the halo using one of two methods, depending 
on the HOD (see Table~\ref{tab:hods}). For some of the HOD models, the satellites are assigned the position and velocity
of dark matter particles, which are randomly chosen from the FOF group belonging to that halo.
For other HOD models, the haloes are positioned randomly, following a \citet{Navarro1997} (NFW)
density profile, using a concentration-mass relation from \citet{Ludlow2014}. 
The satellites are assigned a random virial velocity, where the component in each direction is drawn
from a Gaussian distribution with variance $\sigma^2(M) = G M_\mathrm{vir} / 2R_\mathrm{vir}$.
The virial radius is determined from the halo mass using a relation measured from the MDPL2
simulation \citep{Klypin2016}.

The method used to position satellite quasars affects the clustering measurements
on very small scales (on the order of a few $\hMpc$), which are smaller than the scales used in the analysis.

For each of the 20 HOD models, we generate 100 independent random realizations. Since the quasar duty cycle 
is of the order of 1\%, it is expected that each halo would appear once in the full ensemble (on average).
\highlight{Since the same N-body simulation is used, the 100 realizations of each HOD are not
independent. However, we have measured the cross correlation between pairs of mocks, and find that
each pair of individual realizations is uncorrelated, to within the statistical precision,
due to the low duty cycle.
}

To convert to redshift space, the velocity of each halo is projected along the line of sight of the 
observer. Before converting to redshift space, the periodic box is first replicated, which takes
into account any objects which move in or out of the box. The box is then cut back to the original volume.

The observer is positioned at a distance of $2800~\hMpc$ from the
centre of the box, in the positive or negative direction along one of the simulation axes.
This places the observer so that it is facing one of the 6 box sides. Since the observer
is not at infinity, the lines of sight are not parallel.
The choice of observer affects the clustering measurements, and particularly impacts
the quadrupole and hence the best fit values of $\fsig$ (see Section~\ref{sec:observer_position}). 
This is a statistical effect, which is due to cosmic variance in the finite simulation
box. In order to mitigate the impact of observer position on the $\fsig$ measurements,
we alternate between which of the 6 box sides the observer is placed it. 
As is shown in \citet{Smith2020}, there is an anti-correlation between measurements of the
quadrupole with different lines of sight. Therefore, the reduction in the error
when averaging over all the box sides is greater than what would be gained by tripling
the simulation volume.

\highlight{The comoving position to each QSO is then converted to a cosmological redshift,
$z_\mathrm{cos}$. The observed redshift, which takes the line-of-sight velocity, $v_\mathrm{los}$,
into account, is calculated from $(1+z_\mathrm{obs})=(1+z_\mathrm{cos})(1 + v_\mathrm{los}/c)$,
which adds evolution of velocity (and hence $f$) to the mocks.}
For each mock, to test the effects of redshift smearing and catastrophic redshifts,
we create a `no smearing', `Gaussian smearing', `realistic smearing' and
`castastrophic redshift' version, as described in Section~\ref{sec:observational_effects}.

\subsection{Blind mocks} \label{sec:blind_mocks}

\highlight{In addition to the non-blind mocks, we also create mocks in rescaled cosmologies, which
are analysed blindly, using the OuterRim cosmology as the fiducial cosmology. We describe the
method for rescaling the cosmology of the simulation in Sections~\ref{sec:rescaling_cosmology}-\ref{sec:rescaling_cosmology_power_spectrum}, 
which is validated in Section~\ref{sec:rescaling_cosmology_validation}. In
Section~\ref{sec:blind_mocks_method}, we apply this method to the OuterRim simulation to
construct the blind mocks.}

\subsubsection{Changing the cosmology} \label{sec:rescaling_cosmology}

In order to rescale the halo catalogues, we use the methodology of \citet{Mead2014a,Mead2014b}. 
This is an extension of the method of \citet{Angulo2010}, where the rescaling was
applied to the dark matter particles. Scaling the halo catalogue makes it easier to 
modify the cosmology of large simulations, such as OuterRim, 
since the halo catalogue is much smaller in size, and particle data is not always available.
Recently, an alternative method of warping the simulation cosmology was outlined in \citet{Garrison2019},
which requires a set of N-body simulations in different cosmologies. However, since
there is only one OuterRim simulation, in one cosmology, we use the method 
of \citet{Mead2014a}. In this section, we give a brief overview
of the rescaling procedure. Our Python implementation is publicly
available.\footnote{\url{https://github.com/amjsmith/rescale-cosmology}}

The aim of the rescaling procedure is to take a snapshot of the OuterRim simulation at redshift $z$,
and scale the halo properties to a new `target' cosmology at redshift $z'$.
Quantities in the target cosmology are denoted with a prime, while quantities without a prime
are in the original OuterRim cosmology.
To create our blind mocks, we rescale the simulation to a target cosmology $z'=1.433$, 
which is the same redshift as the unblind mocks, and close to the effective redshift of the quasar sample.

\subsubsection{Global rescaling of simulation units}\label{sec:rescaling_cosmology_scaling}

The first stage aims to modify the halo mass function for the new cosmology. This is done 
with a global rescaling of the units of the simulation. Comoving position vectors, 
$\vect{x}$, in units $\hMpc$ are scaled by a factor $s$,
\begin{equation}
\vect{x}' = s\vect{x},
\end{equation}
which as a result also scales the box size to $L'=sL$. For masses, $M$, 
in units $\hMsun$, the scaling is
\begin{equation}
M' = s_\mathrm{m} M \equiv s^3 \frac{\Omega_\mathrm{m}'}{\Omega_\mathrm{m}} M,
\end{equation}
where $\Omega_\mathrm{m}$ is the matter density parameter.
Finally, velocities $\vect{v}$, in proper $\kms$, are scaled as
\begin{equation}
\vect{v'} = s_\mathrm{v} \vect{v} \equiv s \frac{H'(z')f'(z')}{1+z'} \frac{1+z}{H(z)f(z)} \vect{v} 
\end{equation}
where $H(z)$ is the Hubble parameter and $f(z)$ is the growth rate.


\highlight{
The values of $s$ and $z$ are found such that the rms linear density fluctuation in the rescaled
cosmology, $\sigma(R/s,z)$, matches that of the target cosmology, $\sigma'(R,z')$
\citep[by minimizing equation~2 of][]{Mead2014a}.
}
Simple models of the halo
mass function only depend on $\sigma(R,z)$ \citep[e.g.][]{Press1974,Sheth1999}, so
if $\sigma(R,z)$ is correctly reproduced,
the halo mass function will also be correct. This is not strictly true for the mass
function of a N-body simulation, but deviations in the mass function are expected
to be of the order of a few percent \citep{Mead2014a}.

For many potential target cosmologies, the
value of $z$ found will not correspond to
a simulation snapshot. We therefore only choose
cosmologies where the input redshift, $z$, matches the redshift of one of the OuterRim snapshots. This
places a constraint on the combinations of cosmological parameters that
are allowed. Despite this, it is still possible to rescale the simulation to
a wide variety of new cosmologies.

\subsubsection{Modifying the power spectrum}\label{sec:rescaling_cosmology_power_spectrum}

The scaling of comoving positions by a factor $s$ also shifts the BAO scale to a new position 
$s r_\mathrm{bao}$. However, this is not necessarily the correct BAO scale, $r'_\mathrm{bao}$, 
for the target cosmology.
This can be seen in Fourier space as residual wiggles when taking the ratio of the rescaled power
spectrum to the target power spectrum. In addition, the overall shape of the scaled power
spectrum, while close to the target cosmology, is not necessarily correct.
To correct the shape of the power spectrum, small displacements are applied
to the rescaled positions and velocities of each halo, 
using the Zel'dovich approximation.

The Lagrangian displacement field, $\vect{\psi}$ (see Eq.~\ref{eq:displacement_field}), 
is related to the matter density field, $\delta$. In Fourier space, this can
be written as
\begin{equation}
\vect{\psi_k} = -i \frac{\delta_k}{k^2} \vect{k}.
\label{eq:displacement_field_fourier}
\end{equation}
Therefore, the displacement field can be obtained from the Fourier transform of the
density field. The density field can be determined from the halo catalogue, after the first
stage of rescaling, by computing the 
overdensity of haloes on a grid, and debiasing using the effective bias of the 
sample of haloes. We use a grid with $750^3$ cells, which corresponds to 
a cell size of $\sim 4 \hMpc$ (the exact cell size depends on the size of the rescaled box).

The differential displacement, due to the difference between the original and
target cosmologies is given by
\begin{equation}
\delta \vect{\psi}'_{\vect{k}'} = \left[ \sqrt{\frac{\Delta'^2_\mathrm{lin}(k',z')}{\Delta^2_\mathrm{lin}(sk',z)}} -1 \right] \vect{\psi}'_{\vect{k'}},
\end{equation}
where 
\begin{equation}
\Delta^2_\mathrm{lin}(k,z) = 4\pi \left(\frac{k}{2\pi}\right)^3 P_\mathrm{lin}(k,z)
\label{eq:dimensionless_power_spectrum}
\end{equation}
is the dimensionless linear power spectrum. 
Each halo is then displaced so that the final position vector is
\begin{equation}
\vect{x''} = \vect{x'} + b'(M') \delta \vect{\psi'},
\end{equation}
where $b'(M')$ is the halo bias. The displacement is multiplied by the halo bias to ensure
that the final rescaled snapshot has the correct mass-dependant bias.
A similar differential displacement is also applied to the velocity field 
\highlight{\citep[see equation~32 of][]{Mead2014a}.}

The full halo catalogue is used to compute the displacement field, so that the displacement field is
not noisy. To ensure that the mass-dependent bias is correct, the halo bias $b(M)$ is used when 
displacing the halo positions. This means that different haloes are displaced by different amounts,
which is unphysical. E.g. a massive halo would be displaced a greater distance than a smaller 
satellite halo, and therefore the large halo could `overtake' the satellite. Since the typical
halo displacement is very small\footnote{The distribution of halo displacements peaks
at $\sim 0.2~\hMpc$, with only $\sim 0.04\%$ of haloes being displaced by greater than
$1~\hMpc$}, any effect on the halo clustering would be
on scales much smaller than are used in our analysis.

In the \citet{Mead2014a} method, the halo bias is calculated using the peak background split. However, 
since haloes in the OuterRim simulation are identified as FOF groups with linking
length $b=0.168$, and not the more standard $b=0.2$, the \citet{Sheth1999} halo bias differs with
the halo bias measured from the simulation. We therefore measure the halo bias directly from
the simulation, from the clustering of haloes in the original OuterRim simulation in 5 mass bins 
at 4 different redshifts,
and modify the parameters of the Sheth-Tormen halo bias to match the measurements of the bias.

\subsubsection{Validation}\label{sec:rescaling_cosmology_validation}

To validate the rescaling procedure, we rescale one of the snapshots of the 
Multidark Planck 2 (MDPL2) simulation \citep{Klypin2016}, 
to the cosmology of the Millennium-XXL (MXXL) simulation \citep{Angulo2012}.
This allows us to compare the halo clustering measurements of the rescaled snapshot
to the MXXL simulation which was run in the target cosmology.
These two simulations are chosen since they both have the same halo mass definition
(FOF groups with linking length $b=0.2$), enabling a direct comparison.
This is different to the OuterRim simulation, which uses $b=0.168$.

The MDPL2 simulation has a box size of $1~\hGpc$, in a Planck cosmology
with $\Omega_\mathrm{m}=0.3071$,
$\Omega_\mathrm{b} = 0.0482$, $h=0.6777$, $\sigma_8=0.8228$ and $n_s=0.96$ \citep{Planck2014}. 
The MXXL simulation 
is a $3~\hGpc$ box, in a WMAP1 cosmology with $\Omega_\mathrm{m}=0.25$, $\Omega_\mathrm{b} = 0.045$, 
$h=0.73$, $\sigma_8=0.9$ and $n_s=1$ \citep{Spergel2003}. 
The MDPL2 snapshot at $z=1.425$ is rescaled to $z=1.67$\footnote{The 
rescaled redshift is not arbitrary, and is determined by the initial and target cosmologies},
which is very close to the redshift of one of the MXXL simulation snapshots \highlight{($z=1.63$)}.
Comoving positions, halo masses, and velocities are scaled by the factors
$s=1.049$, $s_\mathrm{m}=0.940$ and $s_\mathrm{v}=0.993$ respectively.

\begin{figure} 
\centering
\includegraphics[width=\linewidth]{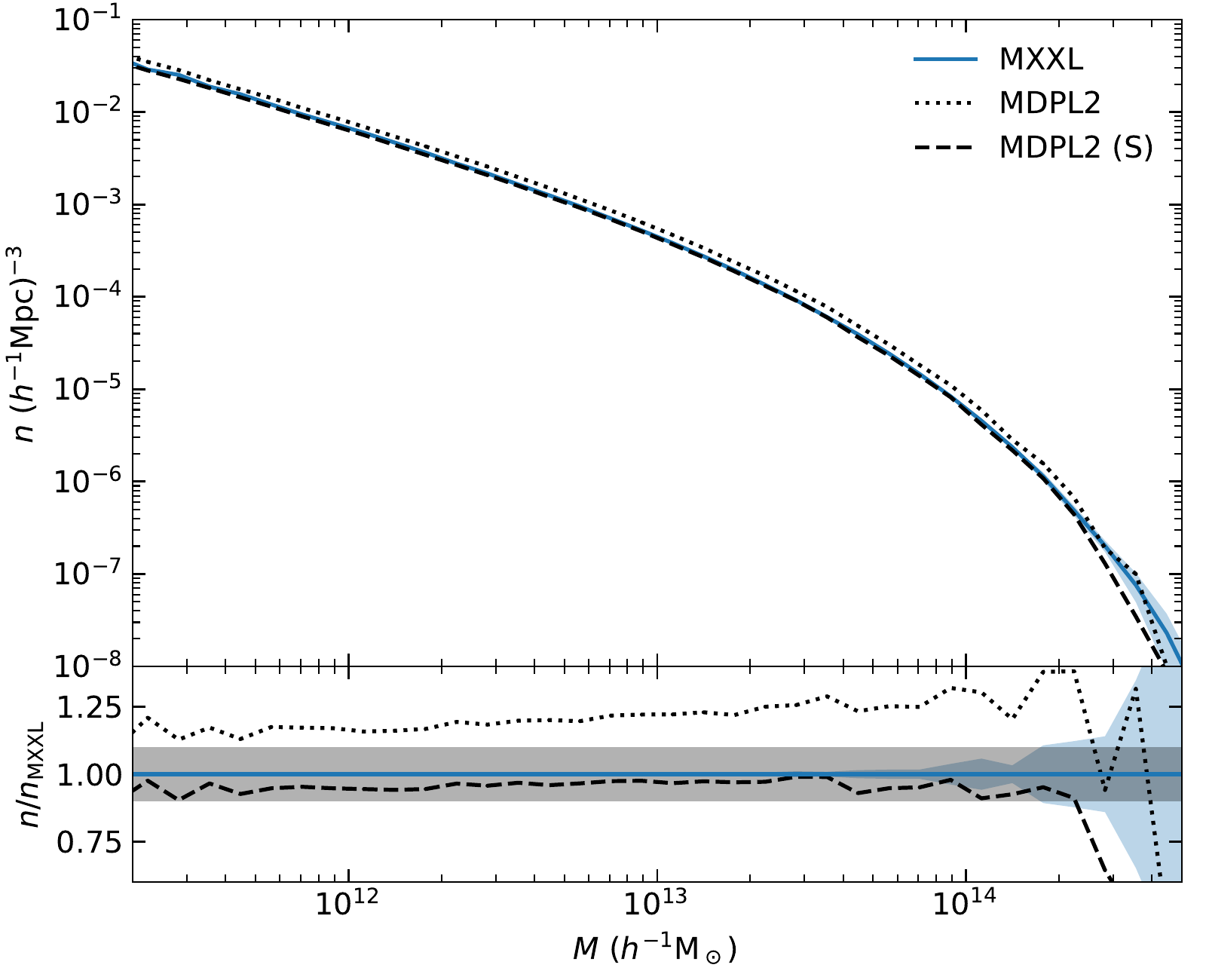}
\caption{\highlight{\textit{Top panel}: Halo mass function of the MDPL2 simulation snapshot at $z=1.425$ before 
(black dotted curve) and after (black dashed curve) rescaling the simulation to the MXXL cosmology at $z=1.67$
The blue curve shows the mass function of the MXXL simulation, where the shaded region indicates the
1$\sigma$ scatter between 8 subvolumes with the same volume as the rescaled MDPL2 snapshot.
\textit{Bottom panel}: Ratio to the MXXL mass function. The grey shaded region indicates 10\%.}}
\label{fig:mass_function_scaled}
\end{figure}

\highlight{
The mass function of the MDPL2 snapshot, after rescaling, is shown in Fig~\ref{fig:mass_function_scaled},
in comparison with the target mass function of the MXXL simulation. The scaling of halo
masses is able to reproduce the target mass function to a level of a few percent, which is in 
agreement with \citet{Mead2014a}. 
}

\begin{figure} 
\centering
\includegraphics[width=\linewidth]{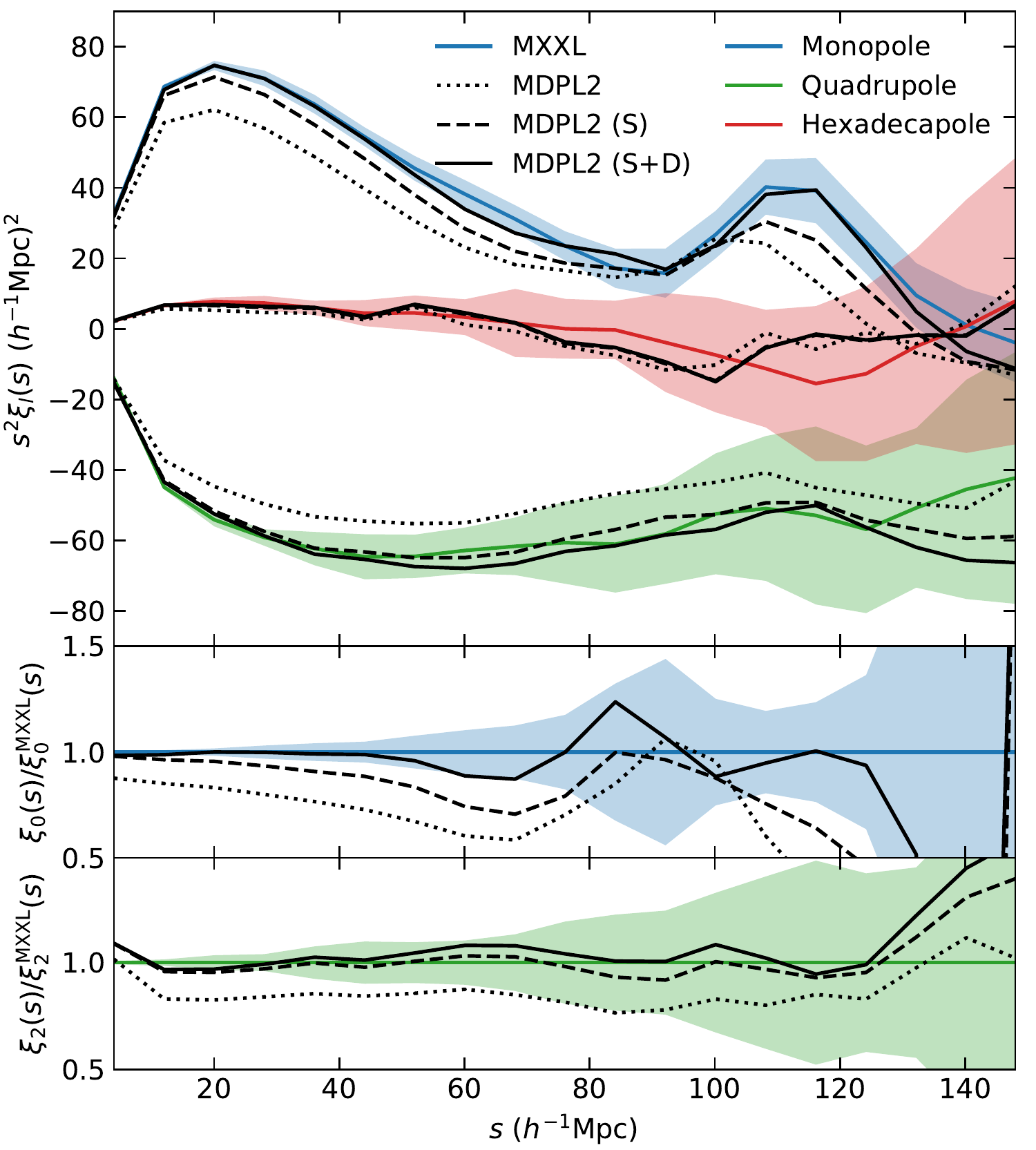}
\caption{\textit{Top panel}: Correlation function monopole (blue), quadrupole (green)
and hexadecapole (red) of the MXXL simulation snapshot at \highlight{$z=1.63$}. 
Black curves show the clustering of the MDPL2 simulation snapshot at $z=1.425$, which
has been rescaled to the Millennium cosmology at $z=1.67$, at each stage of the
rescaling procedure. The dotted curves indicate the clustering in the original 
MDPL2 snapshot, before scaling. The dashed curves show the clustering after
the global scaling of positions, masses and velocities (S). Solid curves indicate
the clustering after scaling, and additional position and velocity displacements (S+D).
For the MXXL simulation, the shaded regions 
indicate the scatter between 8 subvolumes with the same volume as the rescaled MDPL2 
simulation. For both simulations, a mass cut of $10^{12}~\hMsun$ is applied.
\textit{Lower panels}: ratio of the monopole and quadrupole to MXXL.}
\label{fig:xi_rescaled}
\end{figure}

The clustering of the MDPL2 snapshot, measured at each stage of the rescaling procedure,
is shown in Fig.~\ref{fig:xi_rescaled}, in comparison with the MXXL simulation, 
where both catalogues have a mass cut of $M>10^{12}~\hMsun$ applied. For the MDPL2 snapshot,
this mass cut is applied to the rescaled mass, $M'$.
The shaded regions indicate the cosmic variance, estimated from the 1$\sigma$ scatter 
between 8 sub-cubes of MXXL which have the same volume as the rescaled MDPL2 snapshot. Clustering
measurements from MDPL2 are the average of 6 measurements, with the observer placed
at each of the 6 box sides (see Section~\ref{sec:observer_position}). 
In the MDPL2 simulation, the BAO length scale is smaller than for the MXXL simulation, due to the differences
in cosmology.
The first stage of the rescaling procedure shifts the MDPL2 BAO peak 
to larger scales, by the factor $s$, but this is not enough to reproduce the monopole of the MXXL simulation.
The second step of displacing the halo positions is necessary, and brings the
BAO position into excellent agreement with the measurements from the MXXL simulation.
The shape of the monopole on smaller scales is also in good agreement, within the expected
cosmic variance. The amplitude of the quadrupole is also shifted during the rescaling procedure.
Scaling velocities by the factor $s_\mathrm{v}$ has the largest effect on the quadrupole, with a smaller
shift when the velocities are displaced.
After the scaling and displacements, the quadrupole is in good agreement with the MXXL simulation,
and is within the expected cosmic variance.
This validates that the rescaling of halo positions is working as expected. For velocities,
this is less clear, since the cosmic variance is large compared to the shift in the quadrupole
when displacing the velocities.
As an additional check, after creating the blind OuterRim mocks,
we fit the clustering measurements using the CLPT model in the true rescaled cosmology of
the mock (Section~\ref{sec:blind_results_validation}). 
Recovering $\apar=\aperp=1$ and the value of $\fsig$ expected for
that cosmology will validate the position and velocity scaling.

\subsubsection{Methodology for creating blind mocks}\label{sec:blind_mocks_method}

Blind mocks are constructed by applying the rescaling procedure outlined in
Section~\ref{sec:rescaling_cosmology} to snapshots of the OuterRim simulation.
The snapshots at $z=1.494$ and $z=1.376$ are rescaled to 8 new cosmologies at
$z'=1.433$, which is the same redshift as the snapshot used to create the non-blind
mocks. The new cosmologies are chosen by randomly modifying the fiducial OuterRim values of
$\Omega_\mathrm{b}$, $\Omega_\mathrm{cdm}$, $h$ and $\sigma_8$ by multiplying by a random
number generated from a Gaussian distribution. The parameter $n_s$ is then fixed by
requiring the redshift to match the snapshot redshift when rescaling. 
Cosmological parameters are shown in Table~\ref{tab:cosmologies}.

\begin{table*} 
\caption{Cosmologies of the rescaled OuterRim snapshots, which are used to construct the blinded mocks. 
$z_\mathrm{orig}$ is the redshift of the
original snapshot which is rescaled to the new cosmology at $z=1.433$. $s$, $s_\mathrm{m}$ and $s_\mathrm{v}$
are the factors used to scale positions, masses and velocities, respectively. 
Values of $\apar$ and $\aperp$ are calculated using a fiducial cosmology which is the same as
the cosmology of the original OuterRim simulation.
$\fsig$ is evaluated in each of the cosmologies at the target redshift of $z=1.433$.}
\begin{tabular}{ccccccccccccc}
\hline 
 & $\Omega_\mathrm{b}$ & $\Omega_\mathrm{cdm}$ & $h$ & $\sigma_8$ & $n_s$ & $z_\mathrm{orig}$ & $s$ & $s_\mathrm{m}$ & $s_\mathrm{v}$ & $\fsig$ & $\apar$ & $\aperp$ \\
\hline
OR & 0.0448 & 0.2200 & 0.7100 & 0.8000 & 0.9630 & 1.433 & 1.0000 & 1.0000 & 1.0000 & 0.3820 & 1.0000 & 1.0000 \\
\hline
cosmo0   & 0.0461 & 0.2205 & 0.7228 & 0.7742 & 0.9628 & 1.494 & 0.9827 & 0.9555 & 0.9734 & 0.3694 & 0.9988 & 1.0000 \\
cosmo1   & 0.0426 & 0.2360 & 0.6967 & 0.7981 & 0.9384 & 1.494 & 1.0175 & 1.1085 & 1.0308 & 0.3795 & 0.9911 & 1.0002 \\
cosmo2   & 0.0410 & 0.2331 & 0.7405 & 0.7380 & 0.9594 & 1.494 & 0.9087 & 0.7766 & 0.9129 & 0.3514 & 0.9745 & 0.9805 \\
cosmo3   & 0.0447 & 0.2408 & 0.6882 & 0.7903 & 0.9436 & 1.494 & 0.9950 & 1.0624 & 1.0208 & 0.3750 & 0.9966 & 1.0102 \\
cosmo4   & 0.0467 & 0.2202 & 0.6964 & 0.7991 & 0.9815 & 1.376 & 0.9628 & 0.8997 & 0.9788 & 0.3812 & 1.0090 & 1.0105 \\
cosmo5   & 0.0382 & 0.1973 & 0.7197 & 0.8526 & 0.9606 & 1.376 & 1.1049 & 1.1992 & 1.0535 & 0.4103 & 0.9926 & 0.9719 \\
cosmo6   & 0.0541 & 0.2295 & 0.7275 & 0.7910 & 0.9671 & 1.376 & 0.9327 & 0.8688 & 0.9779 & 0.3756 & 1.0113 & 1.0238 \\
cosmo7   & 0.0475 & 0.1844 & 0.7239 & 0.7783 & 1.0280 & 1.376 & 0.9603 & 0.7756 & 0.9086 & 0.3749 & 1.0220 & 0.9981 \\
\hline
\end{tabular}
\label{tab:cosmologies}
\end{table*}

\begin{figure} 
\centering
\includegraphics[width=\linewidth]{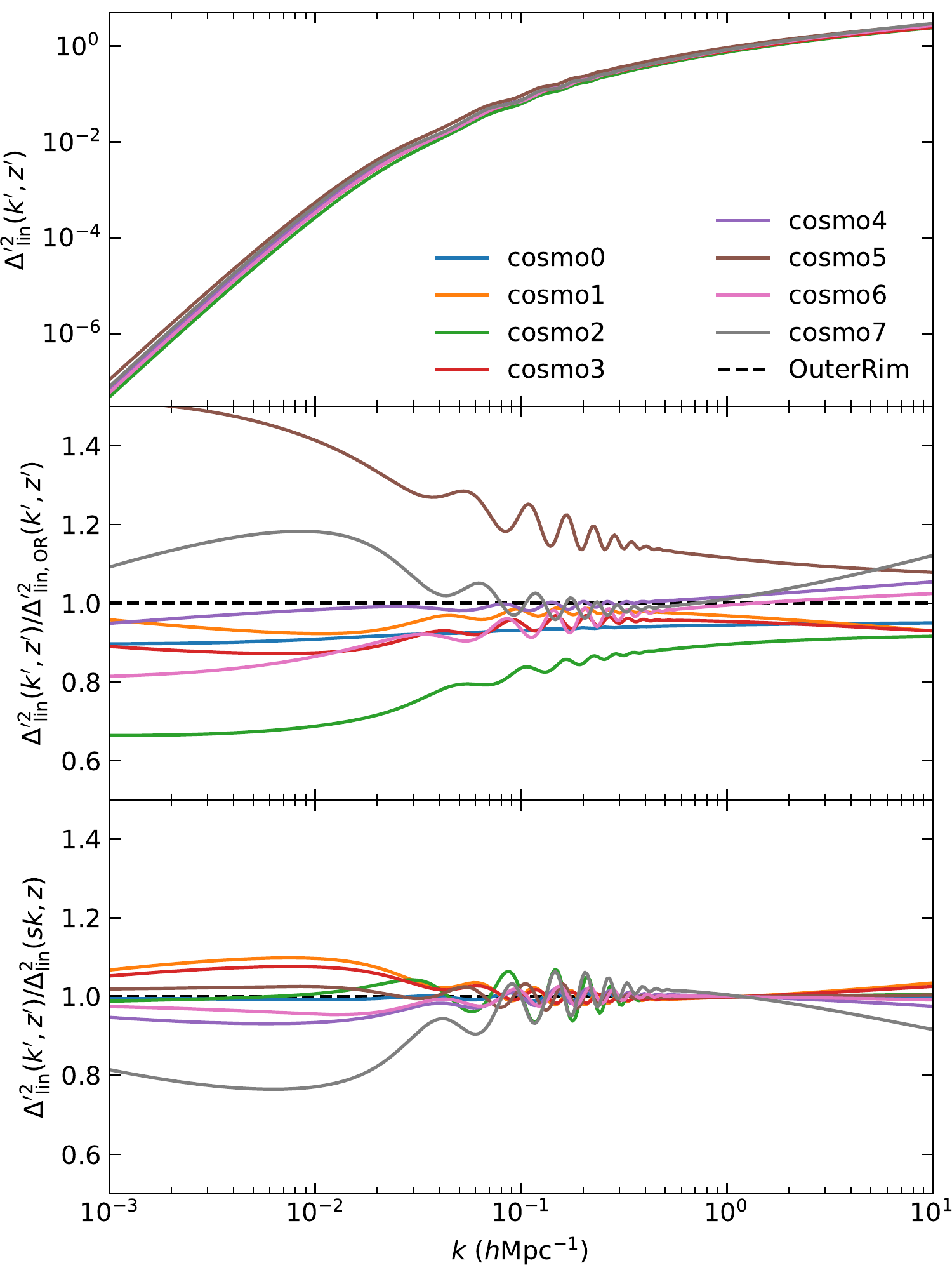}
\caption{\textit{Top panel:} Dimensionless linear power spectra of the 8 new cosmologies at
the target redshift $z'=1.433$ (coloured curves, as indicated in the legend), compared to
OuterRim (black).
\textit{Middle panel:} Ratio of the power spectra to the OuterRim power spectrum.
\textit{Bottom panel:} Ratio of the power spectrum, rescaled by the factor $s$, 
to the target power spectrum.}
\label{fig:pk_rescaled}
\end{figure}

The dimensionless linear power spectrum, $\Delta^2(k)$, defined in 
Eq.~\ref{eq:dimensionless_power_spectrum} is shown in
Fig.~\ref{fig:pk_rescaled} for each of the 8 new cosmologies, compared to the 
original OuterRim cosmology, at $z=1.433$. The second panel shows the ratio
between each of these cosmologies and OuterRim, highlighting the differences
between them. The bottom panel shows the ratio of each of the power spectra, after the initial
rescaling of positions, masses and velocities by the factors $s$, $s_\mathrm{m}$ and
$s_\mathrm{v}$ respectively, to the target cosmology. While this
scaling brings the power spectra close to the target cosmology, there 
are still differences in the shape, and residual BAO wiggles can be seen. This is corrected
for in the second part of the rescaling procedure, where the halo positions 
and velocities are displaced using the Zel'dovich approximation.

The methodology for populating the blind mocks with quasars is the same as for the
unblind mocks. Quasars are added to each halo by re-using some of the HODs, unchanged, 
from Section~\ref{sec:hods}. While the used HODs were tuned to give the same number
density for the non-blind mocks, this is no longer true for the blind mocks, since
the underlying halo mass function changes when the cosmology is changed. The
largest change in number density is by a factor of $\sim 20\%$. 
As before, central quasars are positioned at the centre of each halo, but since particle
data is not available for these snapshots, satellites must be positioned following a
NFW profile. Since the cosmology has been modified, the halo concentrations should
also be modified to reflect this change in cosmology. The positioning of satellites
within their host haloes only affects the clustering on very small scales, which 
are smaller than the scales we consider in the RSD analysis.
This is shown with the non-blind
mocks, where the recovered parameters are unaffected by the method used to position
the satellites. Therefore, we use the same concentration-mass relation as was used
for the non-blind mocks.

For the first part of the blind mock challenge, we create 8 sets of mocks, where each of the
8 rescaled OuterRim boxes is populated using the exact same HOD (\textsc{hod0} from the non-blind
challenge). As before, 100 realizations are generated for each of the cosmologies. Since
the HOD is the same, this can be used to assess the effect of changing the cosmology on
the recovered parameters.

For the second part of the mock challenge, we create 24 sets of mocks. These mocks use
the same 8 rescaled snapshots, which are each populated using 3 different HODs. In some
of these mocks, an additional scaling is applied to the velocities, where all the
velocities are increased or decreased by a few percent. This has the effect of modifying
the expected value of $\fsig$ by the same factor. The cosmology, choice of HOD, and 
velocity scaling factors are given in Table~\ref{tab:blind_mocks}.

\begin{table} 
\caption{Table summarising the second set of blind mocks. Cosmological parameters are
summarised in Table~\ref{tab:cosmologies}, and HOD parameters in Table~\ref{tab:hods}. 
The velocity scaling is an additional scaling applied to velocities,
in which all velocities are scaled by this paramer.} 
\begin{tabular}{cccc}
\hline
Mock & Cosmology & HOD & Velocity scaling \\
\hline
mockb0  & cosmo1 & \textsc{hod}17 & +5\% \\
mockb1  & cosmo4 & \textsc{hod}11 & \\
mockb2  & cosmo0 & \textsc{hod}11 & +5\% \\
mockb3  & cosmo1 & \textsc{hod}11 & \\
mockb4  & cosmo0 & \textsc{hod}17 & \\
mockb5  & cosmo4 & \textsc{hod}2  & \\
mockb6  & cosmo1 & \textsc{hod}2  & -5\% \\
mockb7  & cosmo0 & \textsc{hod}2  & \\
mockb8  & cosmo4 & \textsc{hod}17 & \\
mockb9  & cosmo3 & \textsc{hod}9  & +2\% \\
mockb10 & cosmo7 & \textsc{hod}16 & \\
mockb11 & cosmo5 & \textsc{hod}3  & \\
mockb12 & cosmo7 & \textsc{hod}9  & +3\% \\
mockb13 & cosmo6 & \textsc{hod}9  & -8\% \\
mockb14 & cosmo3 & \textsc{hod}16 & \\
mockb15 & cosmo2 & \textsc{hod}16 & -5\% \\
mockb16 & cosmo2 & \textsc{hod}3  & +3\% \\
mockb17 & cosmo5 & \textsc{hod}9  & +8\% \\
mockb18 & cosmo6 & \textsc{hod}3  & \\
mockb19 & cosmo6 & \textsc{hod}13 & -2\% \\
mockb20 & cosmo3 & \textsc{hod}13 & -3\% \\
mockb21 & cosmo5 & \textsc{hod}13 & \\
mockb22 & cosmo2 & \textsc{hod}13 & \\
mockb23 & cosmo7 & \textsc{hod}3  & \\
\hline
\end{tabular}
\label{tab:blind_mocks}
\end{table}


\section{Non-blind challenge} \label{sec:non_blind_challenge}


\highlight{In this section, we present the analysis and results of the non-blind part of the
mock challenge. These mocks have been constructed using the HOD models and methodology 
described in Section~\ref{sec:non_blind_mocks}.}

\subsection{Clustering measurements}

\highlight{
The correlation function and power spectrum multipoles are measured for each of the individual 
mock catalogues. Correlation functions are measured using the publicly available
correlation function code \textsc{cute} \citep{Alonso2012}, in evenly spaced bins of width
$8~\hMpc$, up to a maximum separation of $168~\hMpc$. A random catalogue is used, which contains
20 times of the number objects that are in the mocks. While the mocks are generated in a cubic box
where $\bar{n}$ is known, we use a random catalogue. This is because the lines of sight are not 
parallel for our choice
of observer position, so periodic boundary conditions cannot be used. For each object in the
random catalogue, a uniform random value is drawn for its $x$, $y$ and $z$ coordinate, in the cubic box.
In all correlation function calculations, an OuterRim fiducual cosmology is used.
Since the volume of the OuterRim box is much larger than the effective volume of the
quasar sample ($V_\mathrm{eff}=1.7~h^{-3}\mathrm{Gpc}^3$), and there are 100 realizations for each HOD, precise measurements
can be made of the clustering and hence the systematic offsets in the cosmological parameter measurements.
Uncertainties in the measurements from each HOD model can be estimated from the scatter between
realizations.}

\begin{figure} 
\centering
\includegraphics[width=\linewidth]{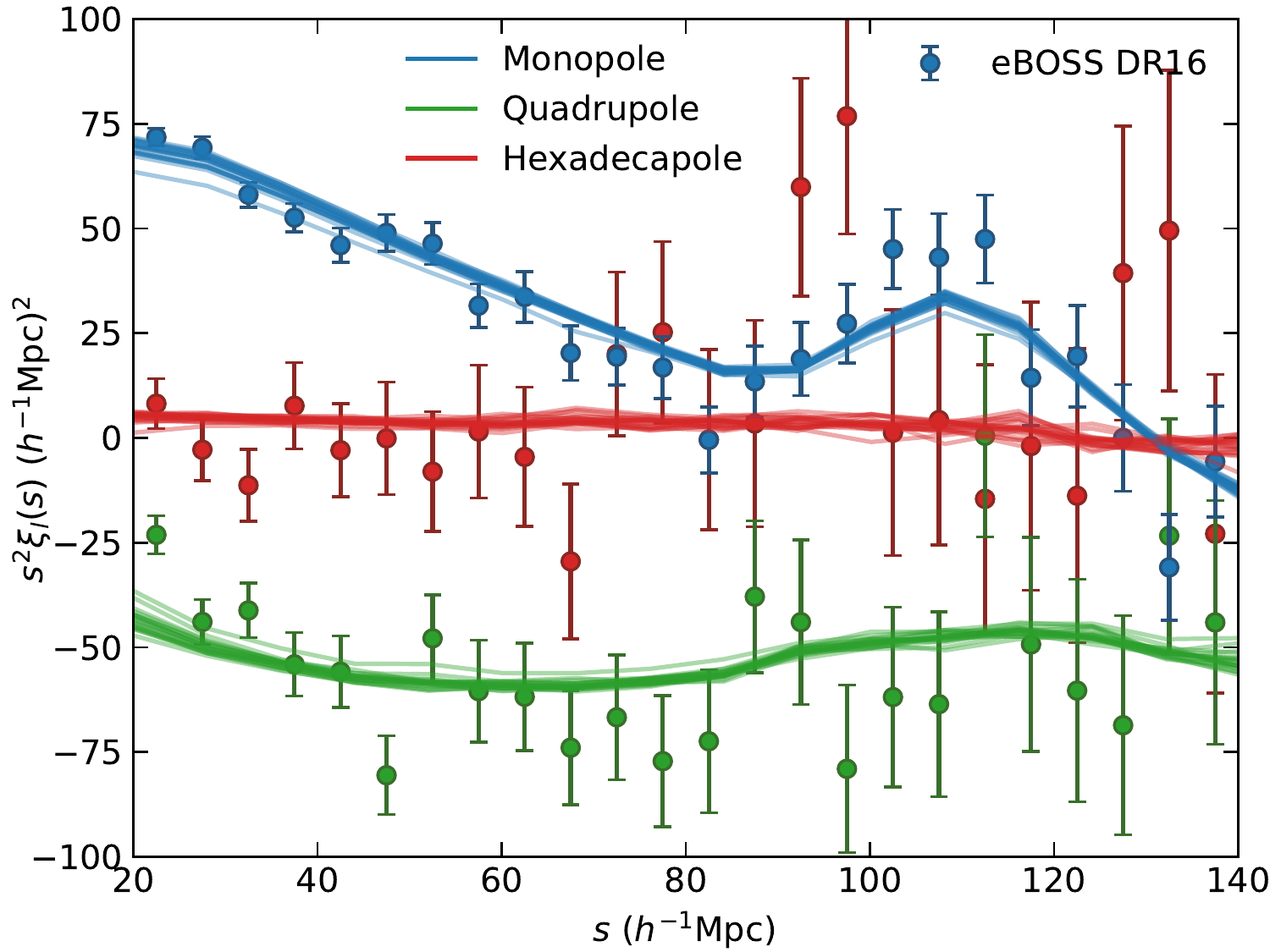}
\caption{The average clustering of each of the 20 sets of non-blind mocks, showing the monopole (blue), 
quadrupole (green) and hexadecapole (red), with no redshift smearing. Points with error bars
show the eBOSS DR16 quasar clustering measurements, \highlight{assuming an OuterRim cosmology, where the
errors are estimated from the EZmocks}.}
\label{fig:xi_nonblind}
\end{figure}

\highlight{
The average correlation function multipoles of the 20 sets of mocks, without redshift smearing, 
are shown in Fig.~\ref{fig:xi_nonblind}, in comparison with the DR16 measurements.
This shows that the clustering of the mocks is in good agreement with the data.
There is a small amount of scatter between the mocks, but this is much
smaller than the DR16 error bars, which are estimated using the EZmocks \citep{Zhao2020}.
}

\highlight{
Power spectra are computed using the \textsc{nbodykit} package \citep{Hand2018}, which uses
the Yamamoto estimator \citep{Yamamoto2006}.
A random catalogue is also used when measuring the power spectrum, 
since the distance to the observer is not infinite, and periodic boundary conditions cannot be assumed.
The computation of the power spectrum is corrected for the effects of the window function
\citep{Wilson2017}, using the implementation of \citet{DeMattia2020}.}


\subsection{Fitting the models} 

The RSD models described in Section~\ref{sec:rsd_models} are fit to the correlation function or power
spectrum multipoles measured from each of the non-blind OuterRim mocks. 
The \Neveux\ model is fit in Fourier space to the power spectrum multipoles, in the
range $0.02 < k < 0.3~\invhMpc$
The covariance matrix used is estimated from the EZmocks \citep{Zhao2020}. Fits are 
done using the \textsc{Minuit} algorithm \citep{iminuit1975}, to minimize $\chi^2$. 

The \Hou\ model is fit in configuration space in 19 evenly
spaced bins of separation, in the range $16<s<168~\hMpc$. 
An analytical Gaussian covariance matrix is used, following the prescription of \citet{Grieb2016}.
Best fit values of each parameter are found from the MCMC chains by taking the median.

CLPT is also fit to the correlation function multipoles, in 16 evenly spaced bins in $s$ 
between $24<s<152~\hMpc$.
The same binning was used as for the \Hou\ model, but with a different minimum and maximum scale.
Fits were done to the mean clustering of each set of mocks, using 
\textsc{Minuit}, with Gaussian covariance matrices.

BAO fits in Fourier space are done in the $k$-range $0.02 < k < 0.23~\invhMpc$, using the
covariance matrix estimated from the EZmocks. The fits are done using \textsc{Minuit}.
BAO fits in configuration space are done with the monopole and quadrupole, using the same
correlation function binning as the \Hou\ model, and a Gaussian covariance matrix.

\subsection{Results} \label{sec:non_blind_results}

\begin{figure*} 
\centering
\includegraphics[width=0.48\linewidth]{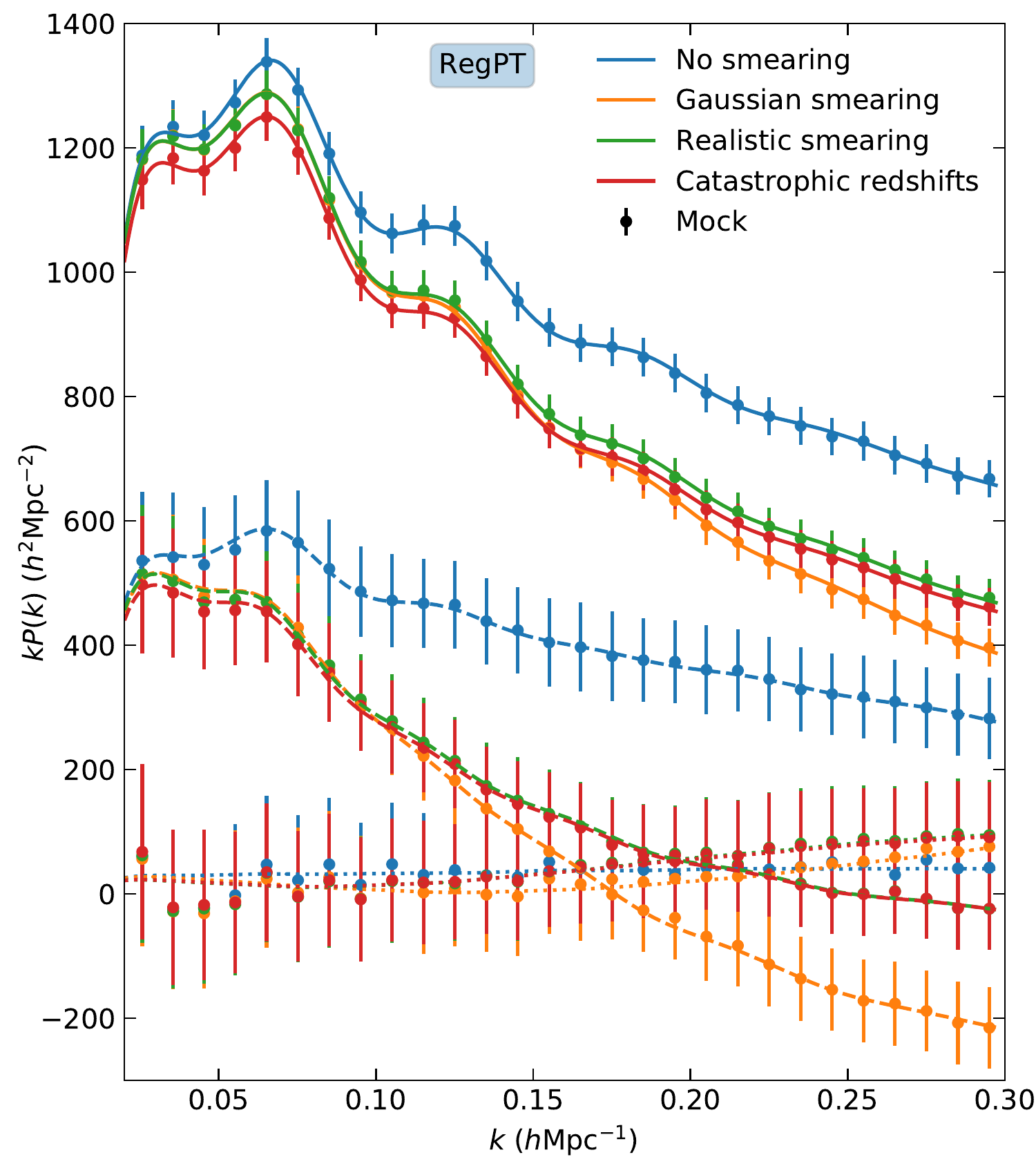}
\includegraphics[width=0.48\linewidth]{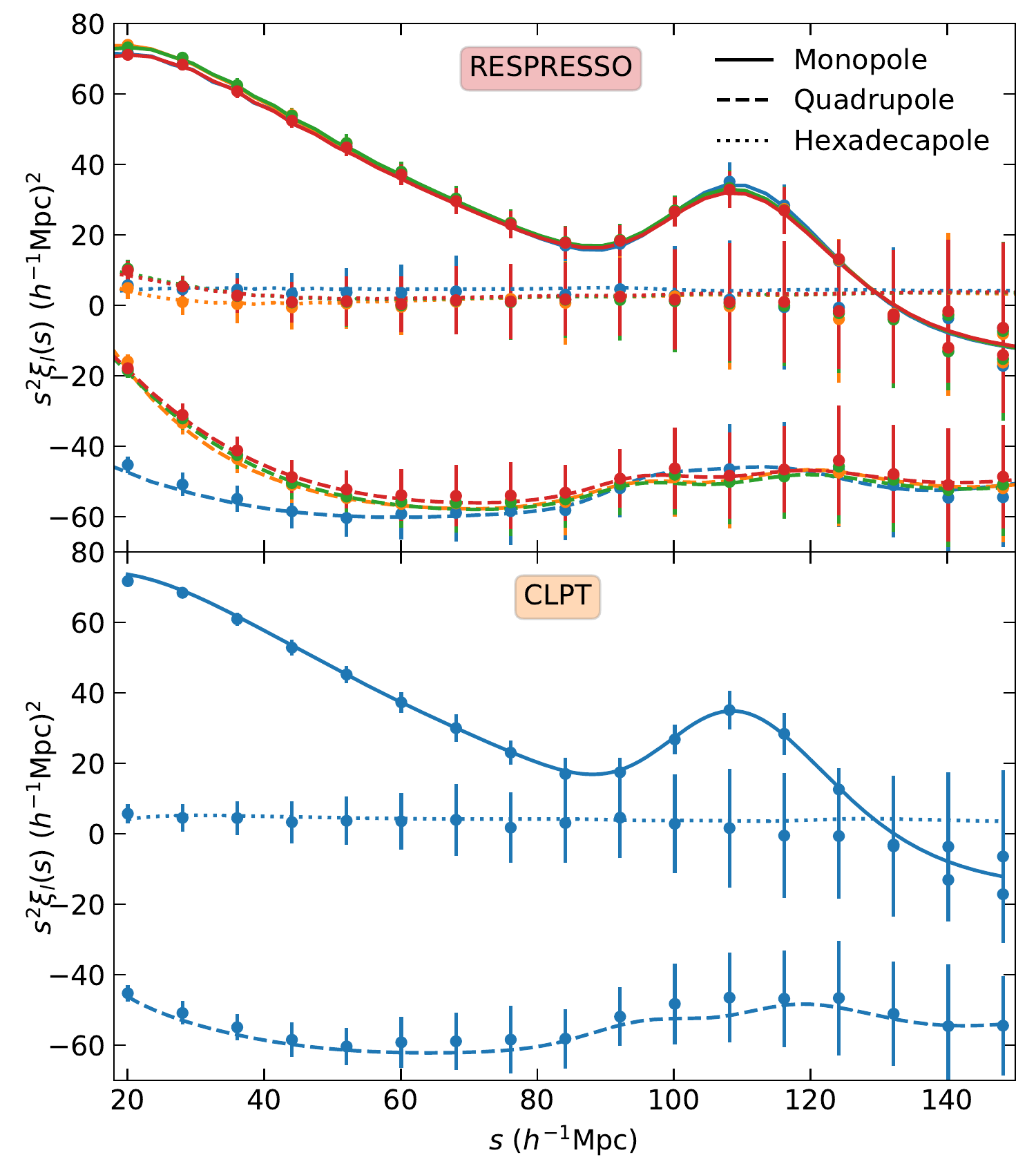}
\caption{\textit{Left panel}: Power spectrum multipoles measured from Mock3, and the best fitting \Neveux\ model, for the
cases of no smearing (blue), Gaussian smearing (yellow), realistic smearing (green) and
catastrophic redshifts (red). Points indicate the average measurement from the 100 mocks,
where the error bar is the 1$\sigma$ scatter. The best fit model is shown by the curves, 
where the solid, dashed and dotted curves are the monopole, quadrupole and hexadecapole,
respectively.
\textit{Right panel}: Correlation function multipoles, and the best fitting \Hou\ model (top)
and CLPT model (bottom).}
\label{fig:model_fits_mock3}
\end{figure*}

Model fits are shown in Fig.~\ref{fig:model_fits_mock3} to the power spectrum and correlation function
measurements of Mock3 (i.e. the mocks constructed using \textsc{hod3}). 
For the \Neveux\ and \Hou\ models, the fits are done with each of the 
different cases of redshift smearing. For the CLPT model, the fit is only done with no redshift smearing.
This figure highlights the impact of redshift smearing on the clustering measurements. 

The impact of redshift smearing on the correlation function is most noticeable in the shape of the quadrupole on small 
scales, but it also has the effect of increasing the amplitude of the monopole on small scales
by a small amount, and also affects the shape of the hexadecapole. 
The difference between the two models of redshift smearing is small, with the most noticeable difference being in the
hexadecapole on small scales. Including catastrophic redshifts reduces the amplitude of the
clustering by a small fraction.

Differences in the power spectrum are much more apparent. Including redshift smearing lowers the amplitude
of the monopole and quadrupole at large $k$, and a larger difference is seen in the shape of the monopole
and quadrupole between the two kinds of redshift smearing.
The two-point clustering statistics measured from the mock are well fit by all models, with
and without redshift smearing.

\begin{landscape}
\begin{figure} 
\centering
\includegraphics[width=\linewidth]{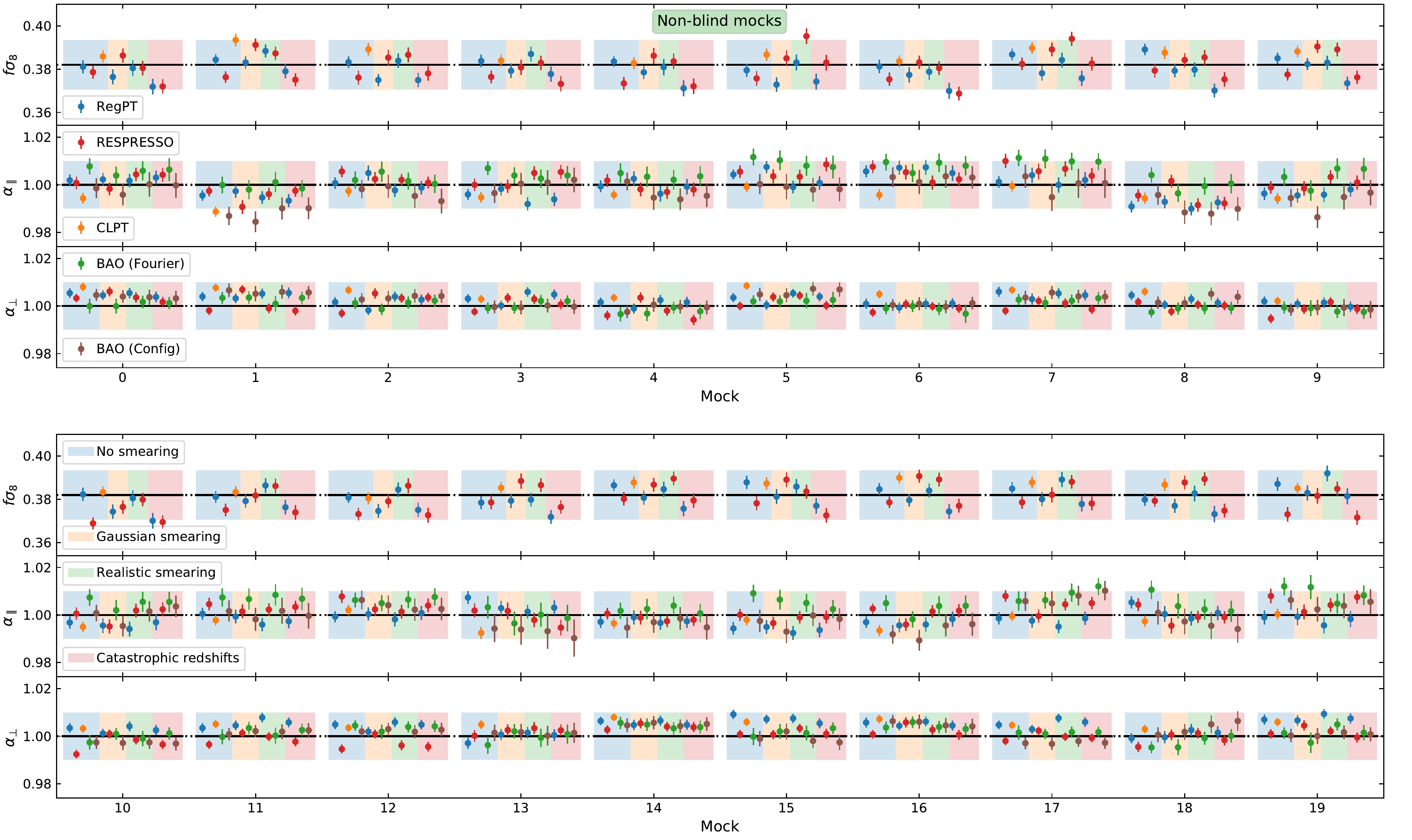}
\caption{Best fitting values of $\fsig$ (upper panels), $\apar$ (middle panels) and $\aperp$ (lower panels),
measured from the 20 sets of 100 non-blind mocks, for all redshift smearing prescriptions.
The results from the \Neveux\ model are shown in blue, \Hou\ in red and CLPT in yellow. 
The BAO fits in Fourier and configuration space are shown in green and brown, respectively.
Points show the mean of the best fit parameters of the 100 mocks, 
and the error bars show the standard error on the mean. 
Shaded regions indicate 3\% for $\fsig$, and 1\% for $\apar$ and $\aperp$. The colour of the shaded region
corresponds to the redshift smearing included in the mocks. No smearing is indicated in blue, Gaussian
smearing in yellow, realistic smearing in green, and realistic smearing with 1.5\% catastrophic redshift errors in red.}
\label{fig:nonblind_results}
\end{figure}
\end{landscape}

The results for all 20 sets of mocks, for all redshift smearing cases are shown in
Fig.~\ref{fig:nonblind_results}. The points in the plot depict the mean values of $\fsig$,
$\apar$ and $\aperp$ measured from each of the 20 sets of 100 mocks, and the error bars indicate the 
error on the mean from the 100 individual mocks. \highlight{The shaded areas indicate 3\% in
$\fsig$ and 1\% in $\apar$ and $\aperp$. This range represents $\sim 30\%$ of the statistical
error, which we consider as the limit that would be acceptable for the systematic error.}

For the case in which there is no redshift smearing (indicated by the blue shading in
Fig.~\ref{fig:nonblind_results}),
the models are able to recover values of $\fsig$ within 3\%, and $\apar$ and $\aperp$
to within 1\% for almost all of the mocks. The results obtained using the \Neveux\ and CLPT models are mostly
in agreement with each other. 
However, there is a small offset between these, and the results from \Hou.
The \Hou\ model, on average, measures values of $\fsig$ and $\aperp$ which are systematically smaller,
by $\sim 2\%$ and $\sim 0.5\%$, respectively, and the values of $\apar$ are $\sim 0.5\%$ larger. 
This difference is explained by the parameter $\sigma_\mathrm{zerr}$ in the FoG prescription used in the
\Hou\ model, which models the redshift error. 
This is a free parameter, but since these mocks do not contain redshift smearing,
fitting the value $\sigma_\mathrm{zerr}$ leads to a small bias in the measurements of 
$\fsig$, $\apar$ and $\aperp$. We tested the effect of fixing $\sigma_\mathrm{zerr}$ to 
$\sigma_\mathrm{zerr}=0$, and this has the effect of increasing the best fit values of $\fsig$ and $\aperp$, 
and reducing $\apar$, bringing the results of all 3 models into agreement.

The results for the case of Gaussian redshift smearing are indicated by the yellow shading
in Fig.~\ref{fig:nonblind_results}.
Again, all the results are within the target ranges for each of the parameters. With Gaussian
smearing, the best fit $\apar$ and $\aperp$ parameters from the \Neveux\ and \Hou\ models are mostly in
close agreement. For $\fsig$ there is a small offset between the two, with the \Hou\ model 
measuring values that are systematically slightly higher by $\sim 2\%$. Since these mocks contain
redshift smearing, which is modelled in the \Hou\ model by the parameter $\sigma_\mathrm{zerr}$,
the best fit parameters measured by the \Hou\ model are closer to the true values, 
compared to the no smearing case.

With the redshift smearing modelled as a more realistic double-Gaussian, the results obtained are
shown by the green shading in Fig.~\ref{fig:nonblind_results}. 
Again, the results are within 3\% for $\fsig$ and 1\% for $\apar$ and $\aperp$ for all models. 
However, now the $\fsig$ is in agreement between the models, with a small difference of $\sim 0.5\%$ in $\apar$. 
Compared to the case of Gaussian smearing, there is very little change
in the results using \Hou. The $\fsig$ from the \Neveux\ model is, on average, $\sim 1\%$ larger,
with a $\sim 0.5\%$ lower value for $\apar$.

Finally, the effect of including catastrophic redshifts is shown by the red shading 
in Fig.~\ref{fig:nonblind_results}. The mocks
here are the same mocks with realistic redshift smearing, but 1.5\% of objects are assigned a random
redshift, from a uniform distribution. This only has a small effect on the $\apar$ and $\aperp$,
but leads to values of $\fsig$ that are systematically lower by $\sim 3\%$. 
This effect on $\fsig$ is expected, since assigning objects
random redshifts will dilute the clustering, reducing the amplitude of the correlation function
quadrupole, resulting in smaller values of $\fsig$.
Despite this shift in $\fsig$, the best fit values are still within 3\% of the expected fiducial value. 

In addition to the full-shape analyses, the BAO-only fits in configuration and Fourier space are also
shown in Fig.~\ref{fig:nonblind_results} for the different redshift smearing cases. For most mocks, the
best fit values of $\apar$ and $\aperp$ are within 1\%. The best fit values of $\aperp$ from the two
BAO models are in good agreement, while there is an offset of $\sim 0.5\%$ in $\apar$, with the 
Fourier space fits measuring values that are systematically larger.
The inclusion of redshift smearing or catastrophic redshifts does not strongly affect the BAO results.

\subsubsection{Observer position}\label{sec:observer_position}

To mitigate the effect of observer position on the average results, we alternate between which of
the 6 box sides the observer is positioned at. If the observer is kept fixed, this can lead to
offsets in the measurements of $\fsig$, $\apar$ and $\aperp$.

\begin{figure} 
\centering
\includegraphics[width=\linewidth]{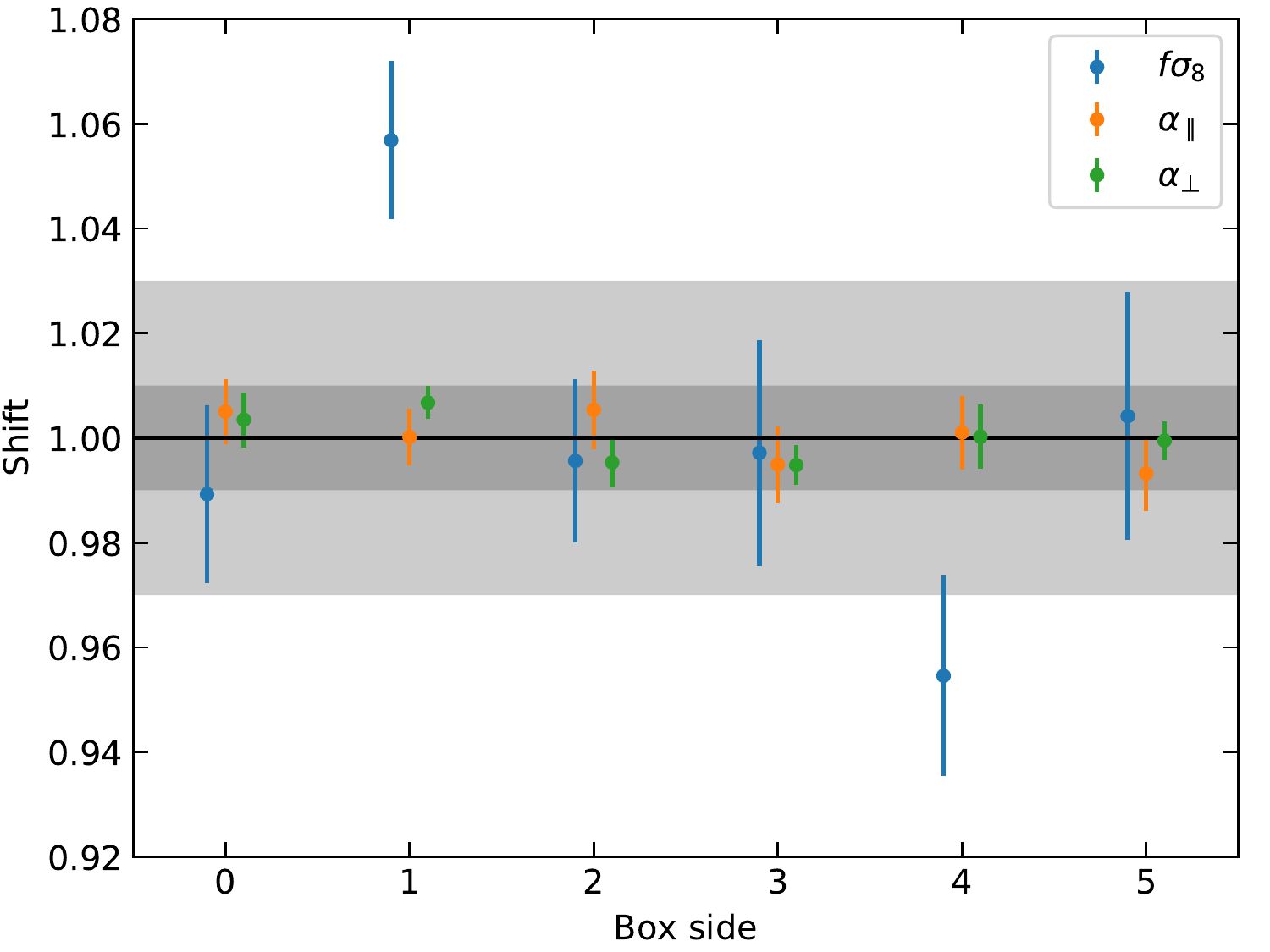}
\caption{Shift in the best fit values of $\fsig$ (blue), $\apar$ (yellow) and 
$\aperp$ (green), measured from Mock0 using the CLPT model, with an observer at each of the 6 box sides,
compared to the average value.
Error bars indicate the error on the mean. Grey shaded regions indicate 1\% and 3\%.}
\label{fig:clpt_fits_box_side}
\end{figure}

The impact of observer position is shown in Fig.~\ref{fig:clpt_fits_box_side} for Mock0, which
shows the offset in the best fit parameters for the subset of mocks with observer at each
box side, compared to the mean of all mocks. For the $\alpha$ parameters, the shifts are small,
and within 1\%. Since the choice of observer position has only a small impact on the monopole,
it is expected that the shifts in the $\alpha$ parameters will be small. However, for
$\fsig$, the shifts can be larger than 5\% for certain observer positions.

\highlight{
The impact of varying the observer position on the quadrupole, and how this propagates through
to the $\fsig$ measurements, is investigated in \citet{Smith2020}.
The offsets seen in the $\fsig$ measurements depend on the box size, and also on the bias of
the tracer. For the QSOs, which are highly biased, large offsets of this magnitude are expected. 
Averaging together the different lines of sight mitigates this effect, reducing the variance
by a factor better than 1/3.}

\subsubsection{Stability of the models}
\label{sec:model_stability}
The results of the non-blind mock challenge show that all models are able to recover the expected value of 
$\fsig$ to within 3\%, and $\apar$ and $\aperp$ to within 1\%, even for mocks that are constructed using
extreme HOD models, which are not motivated by quasar physics. Redshift space distortions in the mock
catalogues are impacted by the choice of HOD, and also the prescription of redshift smearing. 
However, the nuisance parameters which enter our models are able to
absorb these effects in terms of the redshift uncertainty. In this section, we give examples of how the FoG
parameters in the models respond to the HOD (in terms of satellite fraction), and redshift smearing.
We note that the other nuisance parameters also show some sensitivity to these effects.

In the model which we refer to as `\Neveux' (which combines RegPT with the specific RSD 
prescription described in Section~\ref{sec:RegPT}), the free parameters are
$a_\mathrm{vir}$ and $\sigma_v$, while for the `\Hou' model (Section~\ref{sec:respresso}), the
parameters are $a_\mathrm{vir}$ and $\sigma_\mathrm{zerr}$.

\begin{figure}  
\centering
\includegraphics[width=\linewidth]{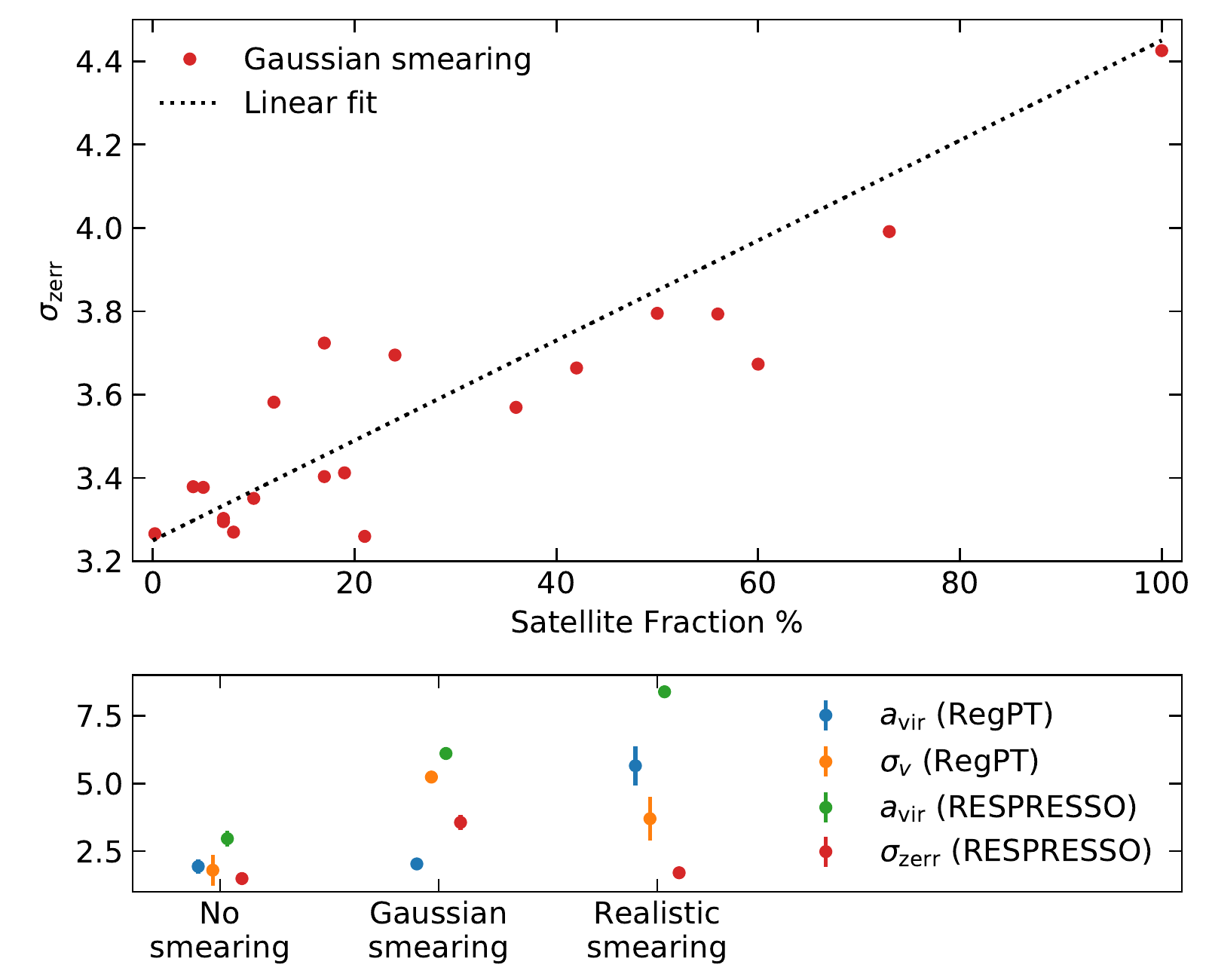}
\caption{\textit{Top panel}: Best fit values of $\sigma_\mathrm{zerr}$ from the RSD prescription used in the 
\Hou\ model, plotted as a function of satellite fraction, for the 20 sets of non-blind mocks with Gaussian 
redshift smearing. The dotted line shows a linear fit.
\textit{Bottom panel}: Average values of the $a_\mathrm{vir}$ (blue) and $\sigma_v$ (yellow) parameters,
from the RSD prescription of the \Neveux\ model, and $a_\mathrm{vir}$ (green) and $\sigma_\mathrm{zerr}$ (red) 
of \Hou\, for the cases of no smearing, Gaussian smearing and realistic smearing.
Error bars indicate the 1$\sigma$ scatter between mocks.}
\label{fig:sig_zerr}
\end{figure}

The different HOD models we consider cover a wide range of satellite fractions, some of which are
very high (e.g. \textsc{hod13} and \textsc{hod8}, where the satellite fractions are 73\% and 100\%, respectively).
The parameters which model the RSDs all show trends with the satellite fraction. 
In the upper panel of Fig.~\ref{fig:sig_zerr} we show the best fit values of 
$\sigma_\mathrm{zerr}$ as a function of satellite fraction, measured by the \Hou\ model, for the mocks
with Gaussian smearing. The values of $\sigma_\mathrm{zerr}$ follow a linear trend, which increases with 
the satellite fraction. Satellite quasars are assigned a random Virial velocity (Section~\ref{sec:non_blind_mocks_method}), 
which is similar to satellites having a larger redshift uncertainty. However, the satallite fraction and redshift
smearing impact the clustering measurements on different scales. The quadrupole is impacted by the satellite fraction
on scales $\sim 40~\hMpc$, while redshift smearing has an impact on larger scales, up to $\sim 90~\hMpc$.

The bottom panel of Fig.~\ref{fig:sig_zerr}, shows the best fit values
of $a_{\rm vir}$ and $\sigma_v$ from the RSD prescription of the \Neveux\ model, and also $a_{\rm vir}$
and $\sigma_\mathrm{zerr}$ of the \Hou\ model, for different cases of redshift smearing. 
The treatment of the FoG factor is very similar between the two models, but the \Hou\ model contains 
a velocity dispersion `offset'.\footnote{In the FoG prescription used in the \Hou\ model, $\sigma_v$ is kept fixed 
to a non-zero value. This results in a $k$ and $\mu$ dependence which does not vanish when $a_\mathrm{vir}=\sigma_\mathrm{zerr}=0$.}
The parameters respond to the redshift smearing schemes, but their behaviours differ between the redshift smearing implementations.
The velocity dispersion offset in the \Hou\ model lowers the values of $\sigma_{\rm zerr}$.
In both models, the parameter $a_\mathrm{vir}$ never vanishes, which demonstrates that the FoG factor
cannot be modelled by a pure Gaussian function. Using a pure Gaussian or a pure Lorentzian FoG factor leads to a 
biased measurement $\fsig$.

We also checked that our results are stable for different choices of binning in the
two-point clustering measurements. For the \Hou\ model, for 5 sets
of mocks, we re-measure the correlation function multipoles in bins of separation $5~\hMpc$, and fit the
model to these correlation function measurements. We find that reducing the bin separation from $8~\hMpc$
to $5~\hMpc$ has a very small effect on our results, and does not change our picture of the 
modelling systematics.

\subsubsection{Non-blind Systematics}

\begin{figure} 
\centering
\includegraphics[width=\linewidth]{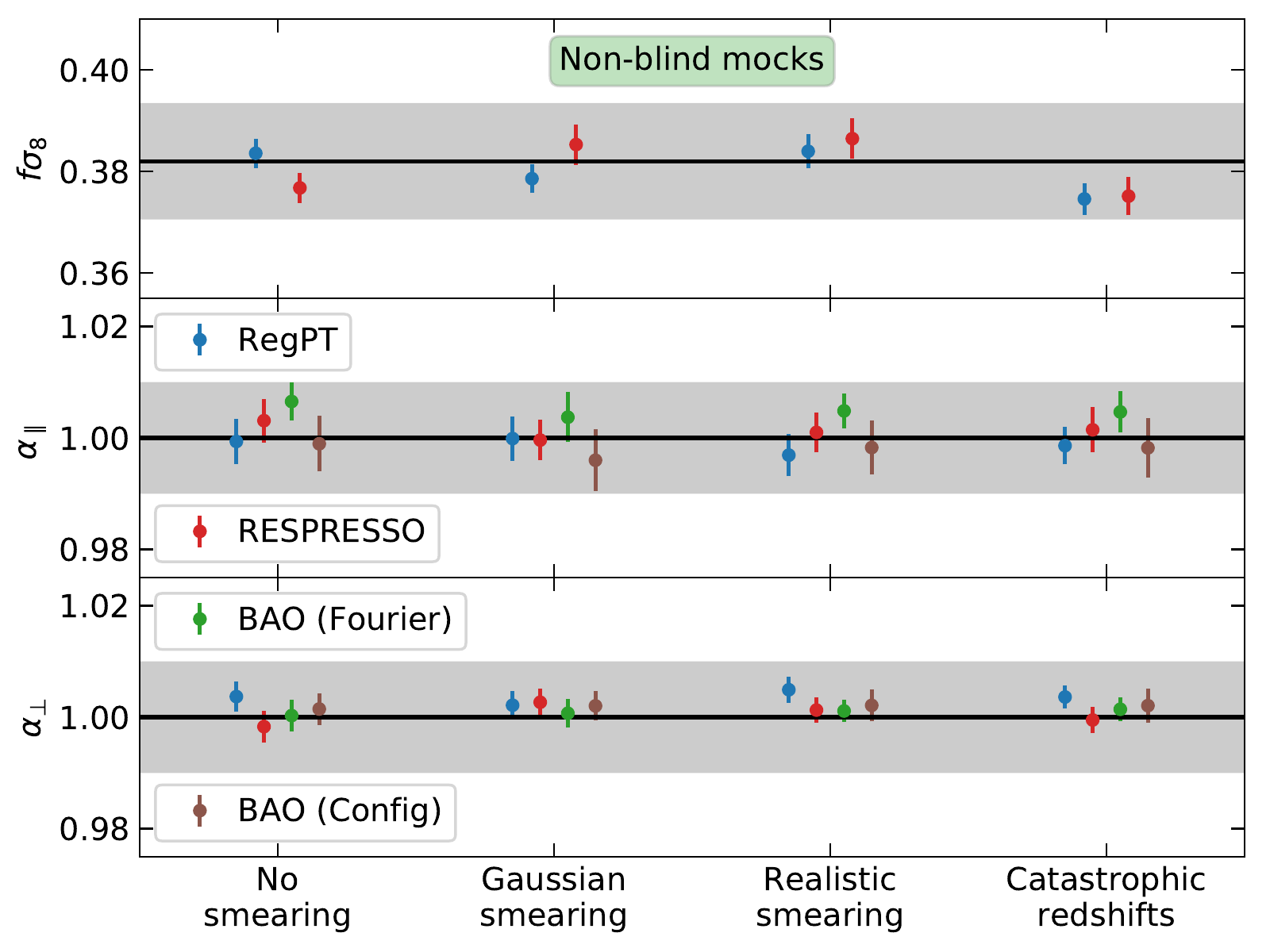}
\caption{Results from the non-blind mock challenge, averaged over all 20 sets of mock catalogues,
for the cases of no redshift smearing, Gaussian smearing, realistic double-Gaussian smearing, and
1.5\% catastrophic redshifts. Points show the average values measured using
\Neveux\ (blue), \Hou\ (red), and BAO fits in Fourier space (green) and configuration space (brown).
where error bars show the 1$\sigma$ scatter between all mocks.
Grey shaded regions indicate 3\% for $\fsig$ and 1\% for $\apar$ and $\aperp$.}
\label{fig:nonblind_summary}
\end{figure}

The results from all the non-blind mocks are summarised in Fig.~\ref{fig:nonblind_summary}. 
The points are the average of all 20 sets of mocks, for each of the different cases of redshift 
smearing, where the error bars show the standard deviation. In order to calculate a 
systematic error, we also calculate the rms, defined as
\begin{equation}
\mathrm{rms} = \sqrt{\sum_{i}^N \frac{(x_i - x_\mathrm{true,i})^2}{N}},
\label{eq:rms}
\end{equation}
where $x_i$ is the value of a parameter measured from mock $i$, $x_\mathrm{true,i}$ is the true 
value, and $N$ is the number of mocks. 
In the case of the non-blind mocks, $x_\mathrm{true,i}$ is constant, and the rms is equivalent
to to adding the offset from the true value, and the standard deviation, in quadrature. 
The average, standard deviation and rms of $\fsig$, $\apar$ and $\aperp$ for the non-blind 
mocks are given in Table~\ref{tab:non_blind_results} for all models.

\begin{table} 
\caption{Results from the non-blind mock challenge using the \Neveux, \Hou, and BAO models, 
averaged over
all 20 sets of mocks, for the cases of no smearing, Gaussian smearing, realistic smearing
and catastrophic redshifts. $\langle \fsig \rangle$ is the mean of the 20 values of $\fsig$,
$\sigma(\fsig)$ is one standard deviation, and $\mathrm{rms}(\fsig)$ is defined 
in Eq.~\ref{eq:rms}. The same values are provided for $\apar$ and $\aperp$}
\begin{tabular}{cccccc}
\hline
\Neveux\ & No & Gaussian & Realistic & Catastrophic\\
& smearing & smearing & smearing & redshifts \\
\hline
$\langle \fsig \rangle$  & 0.3836 & 0.3786 & 0.3840 & 0.3746 \\
$\sigma(\fsig)$          & 0.0029 & 0.0028 & 0.0034 & 0.0031 \\
$\mathrm{rms}(\fsig)$    & 0.0033 & 0.0044 & 0.0039 & 0.0081 \\
\hline
$\langle \apar \rangle$  & 0.9994 & 0.9999 & 0.9969 & 0.9986 \\
$\sigma(\apar)$          & 0.0040 & 0.0040 & 0.0038 & 0.0033 \\
$\mathrm{rms}(\apar)$    & 0.0041 & 0.0040 & 0.0049 & 0.0036 \\
\hline
$\langle \aperp \rangle$ & 1.0037 & 1.0021 & 1.0049 & 1.0036 \\
$\sigma(\aperp)$         & 0.0027 & 0.0025 & 0.0023 & 0.0021 \\
$\mathrm{rms}(\aperp)$   & 0.0046 & 0.0033 & 0.0054 & 0.0042 \\
\hline
\hline
\Hou & No & Gaussian & Realistic & Catastrophic\\
& smearing & smearing & smearing & redshifts \\
\hline
$\langle \fsig \rangle$  & 0.3767 & 0.3853 & 0.3864 & 0.3751 \\
$\sigma(\fsig)$          & 0.0030 & 0.0040 & 0.0040 & 0.0038 \\
$\mathrm{rms}(\fsig)$     & 0.0060 & 0.0052 & 0.0060 & 0.0078 \\
\hline
$\langle \apar \rangle$  & 1.0031 & 0.9996 & 1.0010 & 1.0015 \\
$\sigma(\apar)$          & 0.0039 & 0.0036 & 0.0036 & 0.0040 \\
$\mathrm{rms}(\apar)$     & 0.0050 & 0.0036 & 0.0037 & 0.0043 \\
\hline
$\langle \aperp \rangle$ & 0.9983 & 1.0027 & 1.0013 & 0.9995 \\
$\sigma(\aperp)$         & 0.0029 & 0.0025 & 0.0022 & 0.0024 \\
$\mathrm{rms}(\aperp)$    & 0.0033 & 0.0036 & 0.0026 & 0.0025 \\
\hline
\hline
BAO & No & Gaussian & Realistic & Catastrophic\\
(Fourier) & smearing & smearing & smearing & redshifts \\
\hline
$\langle \apar \rangle$  & 1.0065 & 1.0037 & 1.0048 & 1.0046 \\
$\sigma(\apar)$          & 0.0034 & 0.0045 & 0.0031 & 0.0037 \\
$\mathrm{rms}(\apar)$    & 0.0073 & 0.0058 & 0.0058 & 0.0059 \\
\hline
$\langle \aperp \rangle$ & 1.0003 & 1.0007 & 1.0011 & 1.0014 \\
$\sigma(\aperp)$         & 0.0028 & 0.0026 & 0.0020 & 0.0021 \\
$\mathrm{rms}(\aperp)$   & 0.0028 & 0.0027 & 0.0023 & 0.0025 \\
\hline
\hline
BAO & No & Gaussian & Realistic & Catastrophic\\
(Config.) & smearing & smearing & smearing & redshifts \\
\hline
$\langle \apar \rangle$  & 0.9990 & 0.9960 & 0.9983 & 0.9982 \\
$\sigma(\apar)$          & 0.0050 & 0.0055 & 0.0049 & 0.0053 \\
$\mathrm{rms}(\apar)$    & 0.0051 & 0.0068 & 0.0052 & 0.0056 \\
\hline
$\langle \aperp \rangle$ & 1.0015 & 1.0020 & 1.0021 & 1.0021 \\
$\sigma(\aperp)$         & 0.0028 & 0.0026 & 0.0029 & 0.0030 \\
$\mathrm{rms}(\aperp)$   & 0.0032 & 0.0033 & 0.0036 & 0.0037 \\
\hline
\end{tabular}
\label{tab:non_blind_results}
\end{table}


\section{Blind challenge} \label{sec:blind_challenge}

\subsection{Validation of blind mocks} \label{sec:blind_results_validation}

\begin{figure} 
\centering
\includegraphics[width=\linewidth]{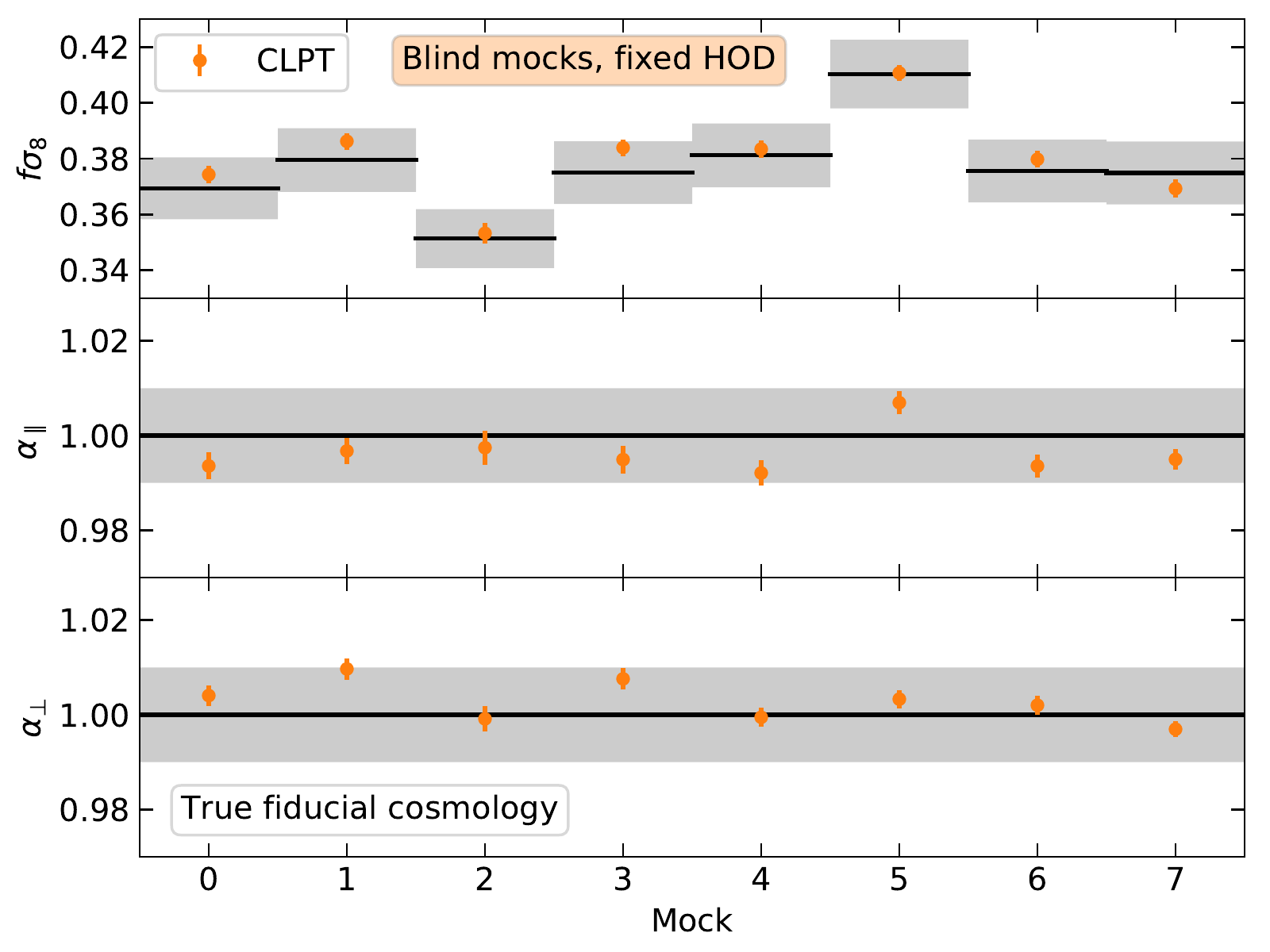}
\caption{Results from analysing the 8 sets of blind mocks, with fixed HOD,
with the CLPT model, using the target cosmology as the fiducial cosmology.
Points with error bars are the results from CLPT, while the true values
are shown in black. The grey shaded regions indicate 3\% for $\fsig$,
and 1\% for $\apar$ and $\aperp$.}
\label{fig:blind_results_rescaling_validation}
\end{figure}

Before analysing the blind mocks in the OuterRim fiducial cosmology, we first analyse the
clustering using CLPT in the target cosmology, to ensure that the rescaling
of the cosmology is correct. This is done with the initial set of 8 mocks, which were 
constructed using the same HOD. Since we use the target cosmology as the fiducial
cosmology, the expected value of $\apar$ and $\aperp$ is 1, while $\fsig$ varies
between the different cosmologies. 
These results are shown in Fig.~\ref{fig:blind_results_rescaling_validation}.
For all the mocks, the measured values
of $\fsig$ are within 3\%, and $\apar$ and $\aperp$ are within 1\%. 
On average, the measured values of $\fsig$ and $\aperp$ are higher than the true values by 
$\sim 1\%$ and $\sim 0.5\%$, respectively,
while $\apar$ is $\sim 0.5\%$ low. The same offsets
are seen when analysing the non-blind mocks using CLPT (Fig.~\ref{fig:nonblind_results}).
This validates that the rescaling method is working as expected,
and we now go on to analyse the mocks blind, using the \Neveux\ and
\Hou\ models, with OuterRim as the fiducial cosmology.

\subsection{Results}

\subsubsection{Blind mocks with fixed HOD} \label{sec:blind_results_fixed_hod}

\begin{figure} 
\centering
\includegraphics[width=\linewidth]{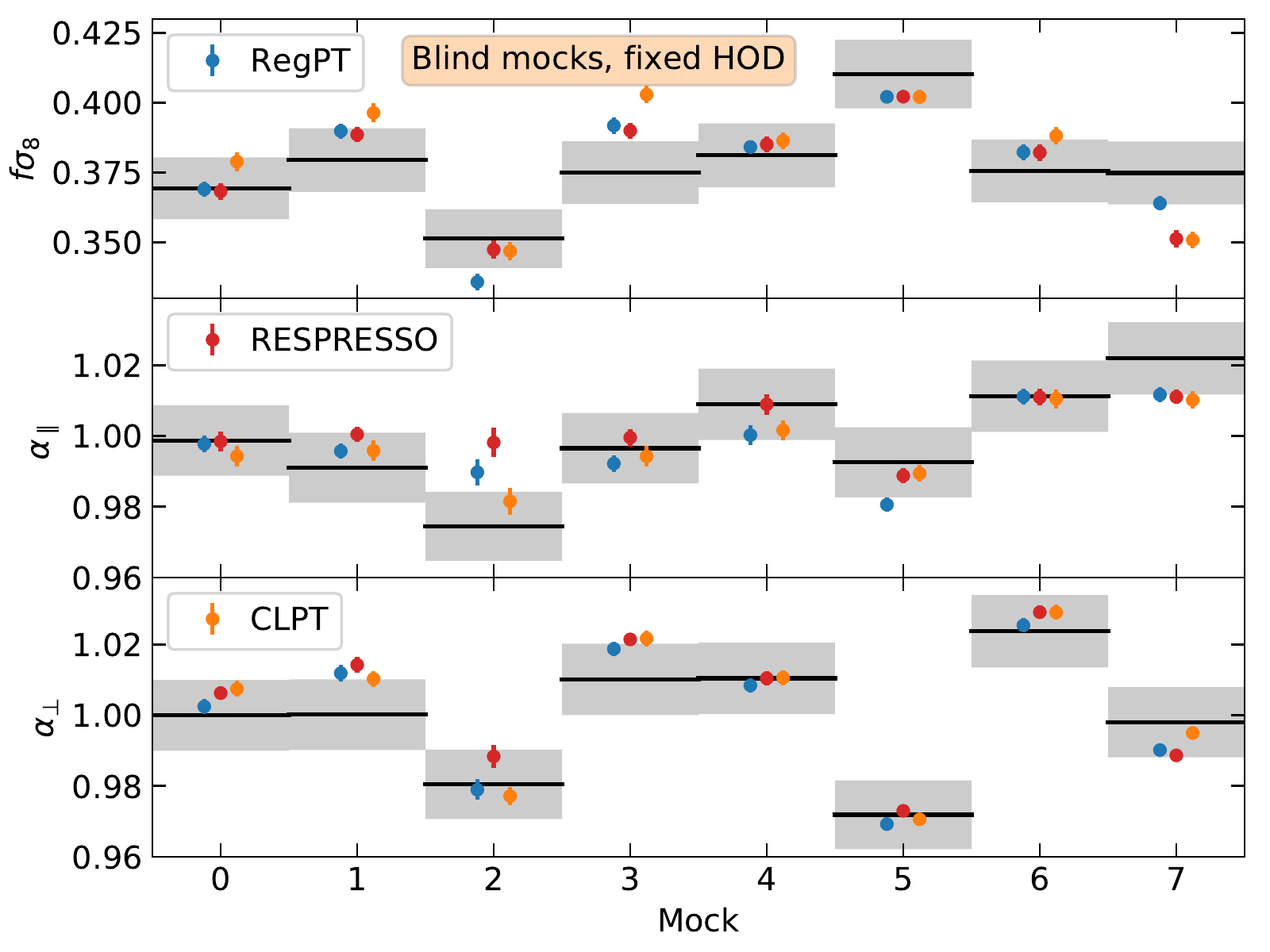}
\caption{Results from the blind mocks, with fixed HOD, using an OuterRim
WMAP7 fiducial cosmology. Points show the results from the \Neveux\ model 
(blue), \Hou\ model (red) and CLPT (yellow), with error bars showing the 1$\sigma$. The expected values are
shown by the black lines, with shaded regions indicating 3\% for $\fsig$ and 1\% for
$\apar$ and $\aperp$.}
\label{fig:blind_results_fixed_hod}
\end{figure}

The results, when analysing the mocks with an OuterRim fiducial cosmology are shown in
Fig~\ref{fig:blind_results_fixed_hod}. Since the fiducial
cosmology no longer matches the true cosmology of the simulation, the expected
values of the $\alpha$ parameters vary from 1 by up to 3\%, but the values of
$\fsig$ are unaffected. The best fit parameters are now in less good agreement with
the expected results than when the fidicual cosmology was the true cosmology of the 
rescaled simulation. For many of the mocks, the measured values are outside the
ranges of 3\% for $\fsig$ and 1\% for the $\alpha$ parameters. For example, the measurements of
$\fsig$ are approximately 5\% too high for mock 3 and 5\% too low for mock 7.
The variation between the measured and expected values is similar for all of the models.

The origin of these differences is not due to the models themselves, since they are able
to recover the expected parameters when the correct fiducial cosmology is used, 
and we have validated that there is no systematic
due to the rescaling procedure. The larger scatter in the results is due to the
fiducial cosmology used.

When the galaxy clustering is measured using an incorrect fidicual cosmology, this
has the effect of shifting the BAO position by the factors $\apar$ and $\aperp$, parallel and perpendicular
to the line of sight, respectively. It has been shown that BAO fitting can recover
unbiased measurements of $\apar$ and $\aperp$, over a range of different choices of
fiducial cosmology \citep[e.g.][]{Vargas-Magana2018,Carter2019}.
However, in order to also measure $\fsig$, it is necessary to perform full shape fits to the correlation 
function and power spectrum multipoles. 
While the position of the BAO peak in the data will be shifted by the expected
$\alpha$ parameters, compared to the template, which is calculated in the fiducial
cosmology, the shape of the correlation function on smaller scales will not
necessarily match the template, leading to biased measurements of
$\fsig$, $\apar$ and $\aperp$. The overall shape of the measured correlation function
doesn't change when a different fiducial cosmology is used, but the shape of the 
template does change.

\begin{figure} 
\centering
\includegraphics[width=\linewidth]{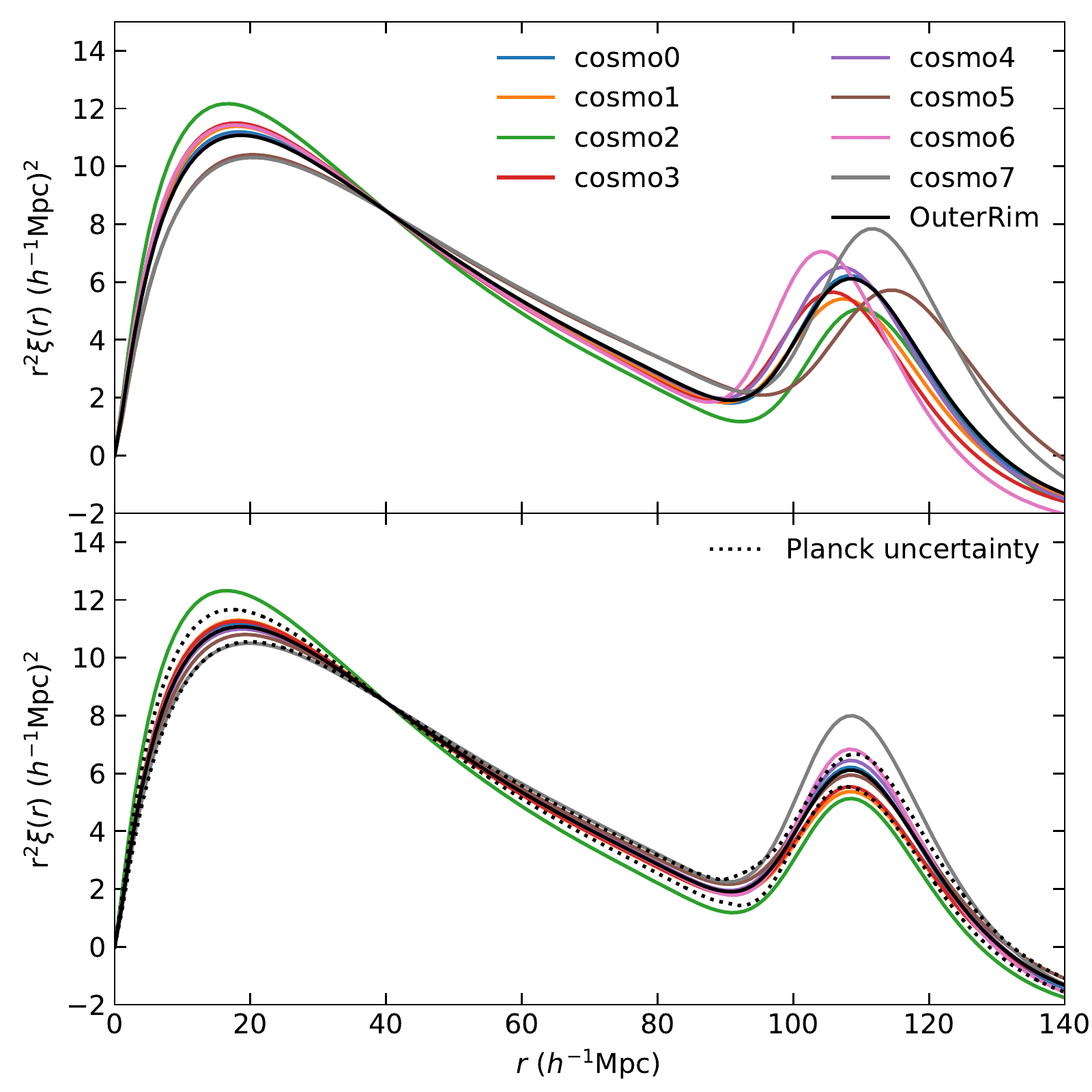}
\caption{\textit{Top panel}: correlation function calculated from the linear power spectrum of the
OuterRim simulation (black dashed curve), and for the 8 new cosmologies (solid coloured curves,
as indicated in the legend). Correlation functions have been shifted vertically so that they
match OuterRim at $r=40~\hMpc$.
\textit{Bottom panel}: same as the top panel, but with a rescaling of $r$ so that the
BAO peaks align. Black dotted lines indicate the maximum change in the OuterRim $\xi(r)$
when varying the cosmological parameters within the 1$\sigma$ errors from Planck.}
\label{fig:xi_shape_comparison}
\end{figure}

Correlation functions are shown in the top panel of Fig.~\ref{fig:xi_shape_comparison} in the cosmologies
of the 8 blind mocks, compared to the OuterRim cosmology. These curves are calculated
from the linear power spectrum, and have been normalised to have the same amplitude 
as OuterRim at $r = 40~\hMpc$ to emphasise the differences in shape. The templates used
in the model fitting also show the same differences in shape. 
In the lower panel of Fig.~\ref{fig:xi_shape_comparison}, the distances have been rescaled to
align the position of the BAO peaks, further emphasising the differences.
For most of the cosmologies, the change in shape, compared to OuterRim, is within the range expected 
when modifying cosmological parameters within the 1$\sigma$ errors of Planck. The most extreme cosmologies are
cosmology 7, which has a high amplitude at the BAO scale, and cosmology 2, where the slope
on small scales is much steeper than OuterRim. If there is a large mis-match in the slope on small scales,
this leads to biased measurements of $\apar$ and $\aperp$. For example, 
it is not possible to rescale the black OuterRim
correlation function curve in Fig.~\ref{fig:xi_shape_comparison} to match both the slope on small
scales, and the BAO peak position, of cosmology 2, so in the best fit model, there will be
a systematic offset in the BAO position.

\begin{figure} 
\centering
\includegraphics[width=\linewidth]{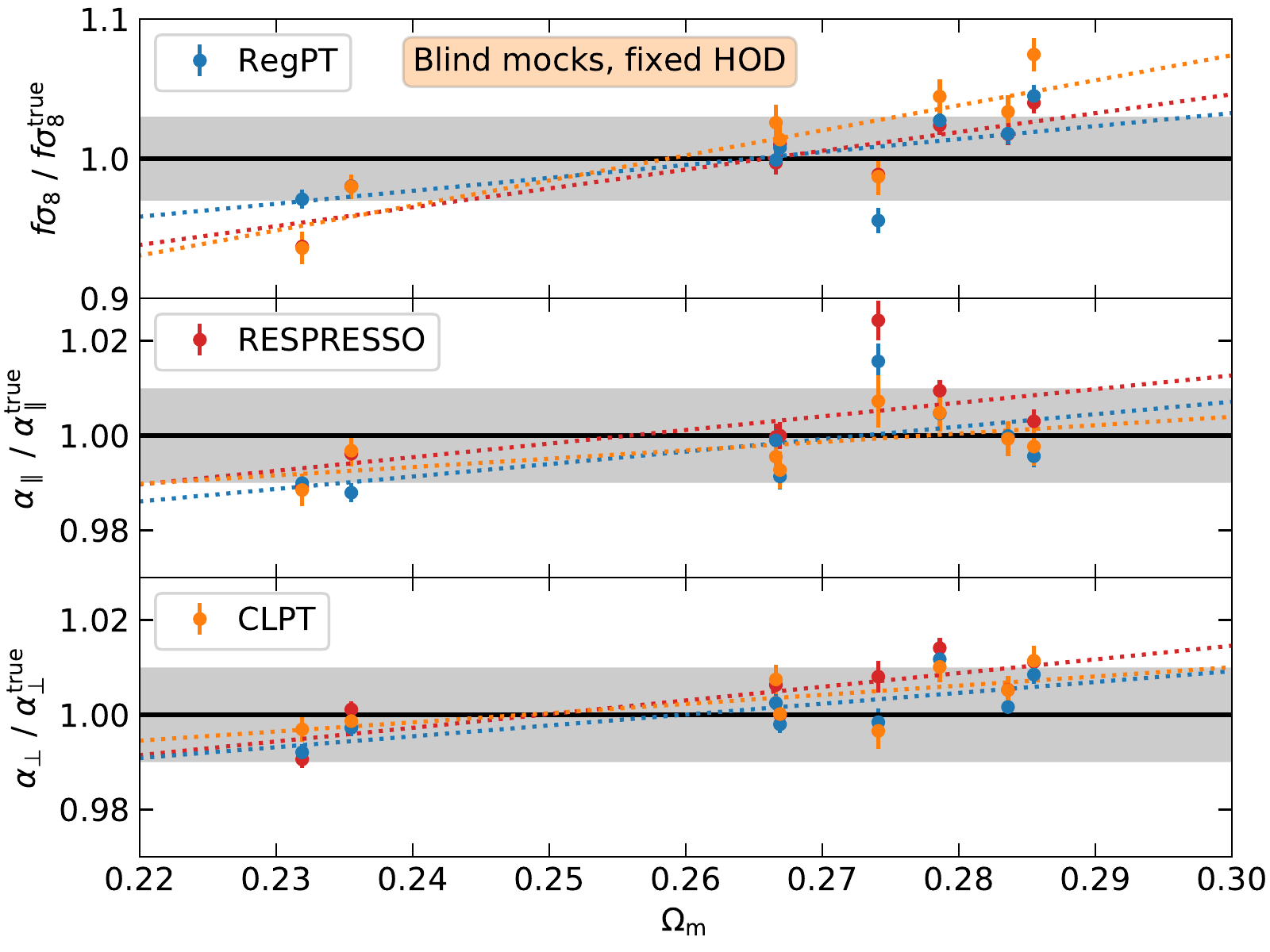}
\caption{Ratio of the measured values of $\fsig$, $\apar$ and $\aperp$ from the blind mocks
with fixed HOD, to the expected value, as a function of $\Omega_\mathrm{m}$. Points
are the results using the \Neveux\ model (blue), \Hou\ model (red) and CLPT (yellow). Dotted
lines show a linear fit to the results from each model, as a function of $\Omega_\mathrm{m}$.
Grey shaded regions indicate 3\% for $\fsig$ and 1\% for the $\alpha$ parameters.}
\label{fig:blind_results_omegam_correction}
\end{figure}

The precise shape of the correlation function is affected by each of the cosmological parameters
chosen (except $\sigma_8$, which only changes the normalisation). The parameter that
has the largest effect on the shape is $\Omega_\mathrm{m}$.
Fig.~\ref{fig:blind_results_omegam_correction} shows the ratio between the measured
and expected values, as a function of $\Omega_\mathrm{m}$, for the three RSD models.
This reveals a correlation with $\Omega_\mathrm{m}$: for cosmologies with low values of
$\Omega_\mathrm{m}$ compared to the fiducial cosmology, 
the best fit values of all the parameters are systematically low, while for the mocks
with larger values of $\Omega_\mathrm{m}$, the measured parameters tend to be higher 
than the expected values. For each of the parameters, the dependence on $\Omega_\mathrm{m}$
is approximately linear for the changes in cosmology considered here, as indicated by the linear fits in 
Fig.~\ref{fig:blind_results_omegam_correction}. The best fit values of
$\fsig$ are within 3\% of these linear fits, and the $\alpha$ parameters 
are within 1\%, with the exception being the mock with cosmology 2. 
This is the cosmology that has the largest difference in the shape of the template, with a steep
slope on small scales (the green curve in Fig.~\ref{fig:xi_shape_comparison}).

\subsubsection{Blind mocks with varying HOD} \label{sec:blind_results_varying_hod}

\begin{figure*} 
\centering
\includegraphics[width=0.9\linewidth]{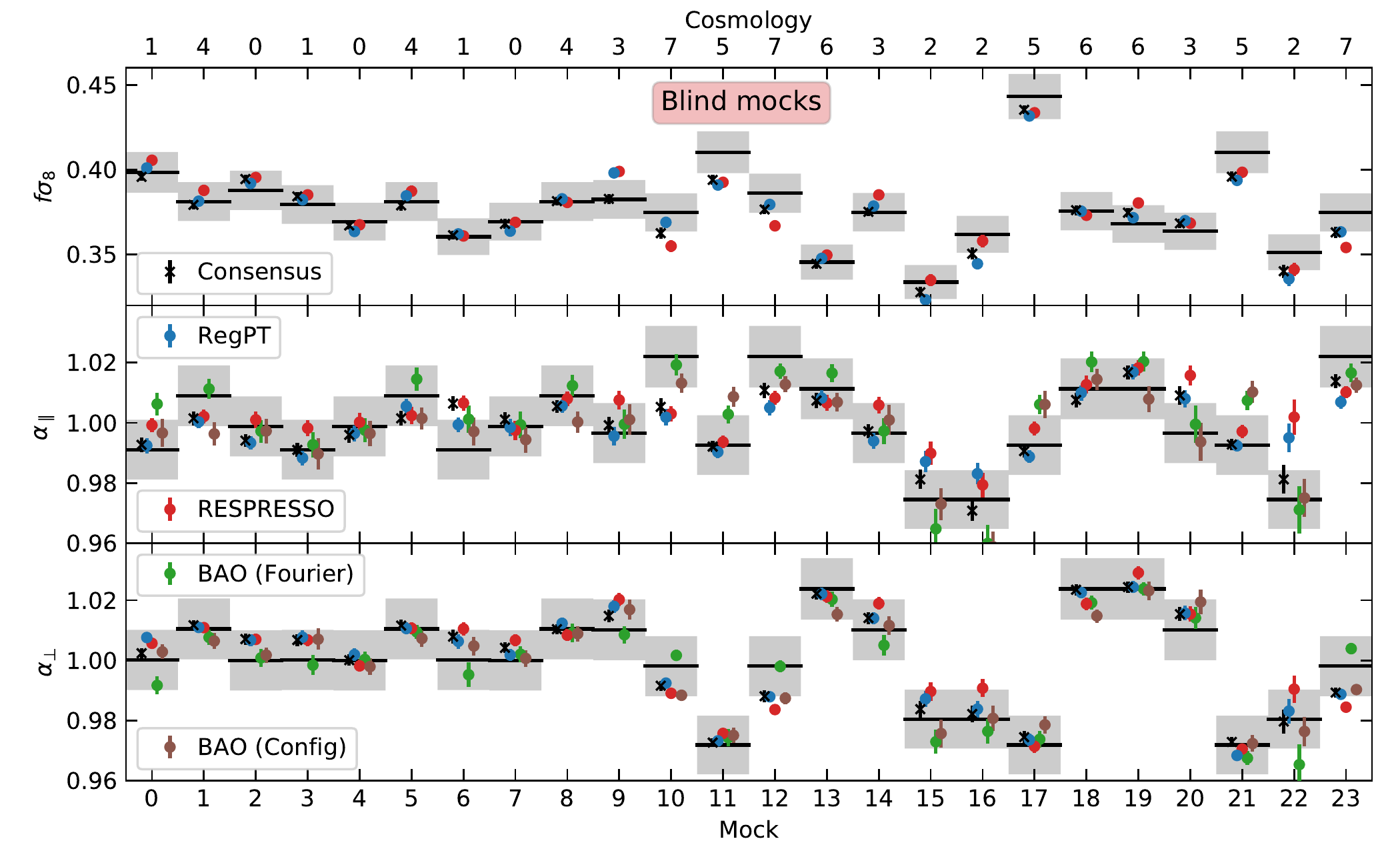}
\caption{Results from the blind mock challenge, for the 24 sets of blind mocks in
which the HOD was varied, and an additional velocity scaling was also applied. 
Points show the average results from the \Neveux\ model (blue), \Hou\ model (red),
combined consensus (black), and BAO fits in Fourier space (green) and configuration
space (brown) of $\fsig$, $\apar$ and $\aperp$. Black lines indicate the true values, with the
shaded regions showing 3\% for $\fsig$ and 1\% for the $\alpha$ parameters.}
\label{fig:blind_results_full}
\end{figure*}

\begin{figure*} 
\centering
\includegraphics[width=0.9\linewidth]{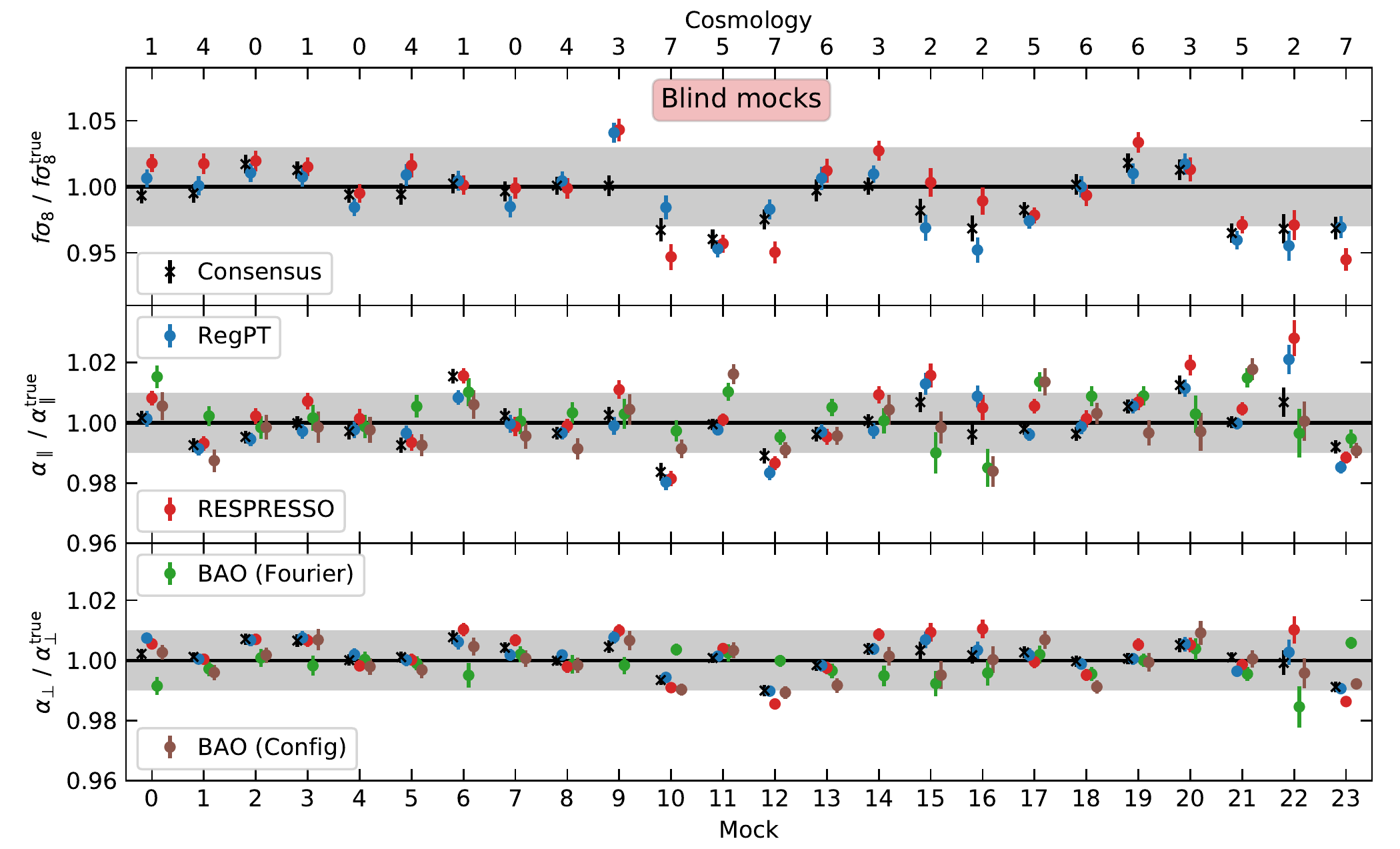}
\caption{Same as Fig.~\ref{fig:blind_results_full}, but showing the ratio to the
true values.}
\label{fig:blind_results_full_ratio}
\end{figure*}

Clustering measurement from the set of 24 blind mocks (with varying HOD and additional velocity scaling) are fit using the \Neveux\ 
and \Hou\ models. Results are shown in Fig.~\ref{fig:blind_results_full}, with the ratios
to the true values shown in Fig.~\ref{fig:blind_results_full_ratio}. For most of the mocks,
the measured value is within 3\% for $\fsig$ and 1\% for $\apar$, but there are some outliers,
which come from the mocks with the largest changes in cosmology (cosmology 2 and 7, 
see Tables~\ref{tab:cosmologies}~\&~\ref{tab:blind_mocks}), and hence have the largest
difference in the shape of the template (Fig.~\ref{fig:xi_shape_comparison}).

The mean, standard deviation and rms from this set of blind mocks is shown in
Table~\ref{tab:blind_results} for all models. There is a small offset
in the best fit values from the \Hou\ model: 
$\fsig$ is systematically $\sim 2\%$ low, and values of $\apar$ are 0.7\% high,
when compared to the true values for each of the cosmologies. This is because
these mocks do not contain redshift smearing, while the parameter $\sigma_\mathrm{zerr}$,
which models the redshift smearing, is a free parameter in the \Hou\ model. As was
shown in the case of the non-blind mocks with no smearing, this leads to a systematic 
offset in the measurements of these parameters. 

The BAO fits are less affected by the assumption of fiducial cosmology.
Compared to the full shape fits, the scatter in the ratios of $\apar$
between the 24 mocks is smaller, leading to smaller values of the rms
in Table~\ref{tab:blind_results}. For $\aperp$, the rms is comparable
to the full shape fits.

\begin{figure} 
\centering
\includegraphics[width=\linewidth]{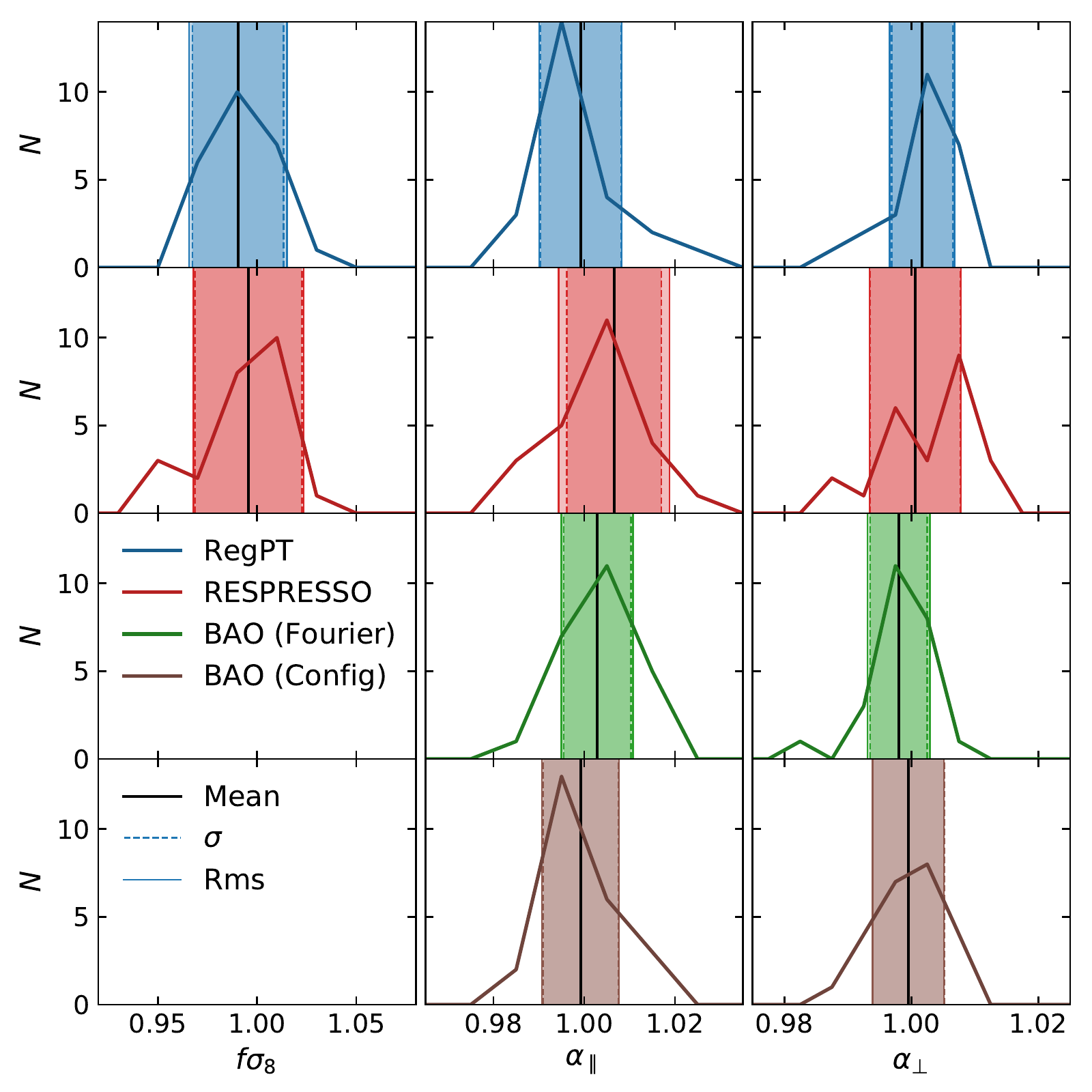}
\caption{\highlight{Distributions of the measured values of $\fsig$ (left column), $\apar$ (middle column) and $\aperp$ (right column) from the 20 sets of blind mocks, for the \Neveux\ model (blue), \Hou\ model (red) and the BAO fits in real space (green) and Fourier space (brown). The vertical lines indicate the mean (black solid line), $\sigma$ (coloured dotted line) and rms (coloured solid line).}}
\label{fig:blind_results_distributions}
\end{figure}

\highlight{The distributions of the measurements of $\fsig$, $\apar$ and $\aperp$
from the blind mocks in Fig.~\ref{fig:blind_results_full_ratio} are shown in
Fig.~\ref{fig:blind_results_distributions}. The mean, standard deviation and rms are indicated by
the vertical lines (for $\fsig$, the rms is divided by the average true value of $\fsig$ from all the
blind mocks). For models where the mean value is close to 1, the standard deviation and rms are 
very close. The value of the rms is primarily driven by the standard deviation, but is slightly 
increased for models with the largest offsets.}

\begin{table*} 
\caption{Results from the blind mock challenge for the \Neveux\ and \Hou\ models, and BAO fits in
Fourier and configuration space, averaged over the 24 blind mocks. For
each of the parameters $\fsig$, $\apar$ and $\aperp$, we calculate the mean, standard deviation
and rms of the ratios to the true values. In the third column, the results of the
non-blind mocks have been used to correct the systematic offsets in the \Hou\ model
due to having the parameter $\sigma_\mathrm{zerr}$ free.}
\begin{tabular}{cccccc}
\hline
& \Neveux & \Hou & \Hou & BAO & BAO \\
& & & (corrected) & (Fourier) & (Config.) \\
\hline
$\langle \fsig/\fsig^\mathrm{true} \rangle$  & 0.9904 & 0.9820 & 0.9957 & - & - \\
$\sigma(\fsig/\fsig^\mathrm{true})$          & 0.0230 & 0.0265 & 0.0269 & - & - \\
$\mathrm{rms(\fsig})$                        & 0.0093 & 0.0123 & 0.0105 & - & - \\
\hline
$\langle \apar/\apar^\mathrm{true} \rangle$  & 0.9992 & 1.0066 & 1.0035 & 1.0029 & 0.9992 \\
$\sigma(\apar/\apar^\mathrm{true})$          & 0.0089 & 0.0104 & 0.0103 & 0.0074 & 0.0083 \\
$\mathrm{rms(\apar})$                        & 0.0091 & 0.0122 & 0.0109 & 0.0079 & 0.0085 \\
\hline
$\langle \aperp/\aperp^\mathrm{true} \rangle$ & 1.0017 & 1.0006 & 1.0022 & 0.9980 & 0.9995 \\
$\sigma(\aperp/\aperp^\mathrm{true})$         & 0.0048 & 0.0071 & 0.0071 & 0.0045 & 0.0057 \\
$\mathrm{rms(\aperp})$                        & 0.0051 & 0.0072 & 0.0072 & 0.0049 & 0.0056 \\
\hline
\end{tabular}
\label{tab:blind_results}
\end{table*}

\subsection{Systematic errors}

The aim of the mock challenge is to estimate the systematic error in the
measured values of $\fsig$, $\apar$ and $\aperp$ in the analysis of the 
eBOSS QSO clustering sample, using the RSD models of \Hou\ and \Neveux.
This systematic error needs to encompass the effects of different HODs,
redshift smearing, catastrophic redshifts, and an incorrect assumption for
the fiducial cosmology.
To calculate the total systematic error, we therefore combine the results of 
the set of 20 catastrophic redshift
non-blind mocks, with the set of 24 blind mocks. 

For the blind mocks, in order to estimate the systematic from different 
cosmologies, we calculate the rms of the 24 sets of mocks (defined in Eq.~\ref{eq:rms}).
However, the \Hou\ model is affected by an additional systematic from having the parameter
$\sigma_\mathrm{zerr}$ free when the mocks do not contain redshift smearing.\footnote{To test the flexibility of the model, $\sigma_\mathrm{zerr}$ is left as a free parameter during the blind analysis. Afterwards, we compared the difference between varying or fixing $\sigma_\mathrm{zerr}$, and we found systematic shifts of $\sim 1\%$, $\sim 0.3\%$ and $\sim 0.2\%$ in $\fsig$, $\apar$ and $\aperp$, respectively.} To remove
this systematic offset, and to measure only the systematic due to the fiducial cosmology, we use the
results from the non-blind mocks with no redshift smearing to apply a correction. 
For $\fsig$, the corrected value is 
\begin{equation}
\fsig^\mathrm{corr} = \left( \frac{\fsig^\mathrm{nb,true}}{\fsig^\mathrm{nb}} \right) \fsig,
\end{equation}
where $\fsig^\mathrm{nb}$ is the average value of $\fsig$ measured from the non-blind mocks
(given in Table~\ref{tab:non_blind_results}), and $\fsig^\mathrm{nb,true}=0.382$ is the true
value of $\fsig$ in the non-blind mocks. 
A similar correction is made to $\apar$ and $\aperp$. The \Neveux\ model is unaffected,
by this, so no correction is needed.
The results, after the correction, are given in the third column of Table~\ref{tab:blind_results}.
The values of rms are slightly larger for \Hou\ than the \Neveux\ model,
but both are at a level of $\sim 0.01$ for $\fsig$ (which is $\sim 2.5\%$ for a value of $\fsig=0.382$), $\sim 1\%$ for
$\apar$ and $\sim 0.6\%$ for $\aperp$. Since the BAO fits are affected less by the
fiducial cosmology, the rms values of $\apar$ and $\aperp$ are smaller than
for the full shape fits.

To measure the effect of different HODs, redshift smearing, and catastrophic redshifts,
we calculate the rms from the `catastrophic redshift' set of non-blind mocks, which include
all these effects. These values are given in the final column of 
Table~\ref{tab:non_blind_results} for each model.
Compared to the blind mocks, the rms values from the non-blind mocks are all smaller.

The total error is calculated by adding the rms from the blind and non-blind mocks in quadrature. 
These are given in Table~\ref{tab:systematic_errors}.
These are conservative estimates for the total systematic errors.
Firstly, adding the errors in quadrature assumes that they are uncorrelated, which is not
necessarily true, and results in a slight overestimate of the total error. 
Also, the blind mocks cover a wide range of different cosmologies. For the non-blind
mocks, the scatter between the many HOD has the largest effect on the measured
rms, while for the blind mocks, the main effect is from the different cosmologies.
However, there is also an additional effect in the blind mocks from changing
the HODs.

The \Neveux\ and \Hou\ models both perform well, with similar systematic errors.
For both RSD models, we take conservative systematic errors of $\sigma_{\fsig}=0.013$, 
$\sigma_{\apar}=0.012$ and $\sigma_{\aperp}=0.008$. These errors are small compared to the errors 
in the data, and are, at most, expected to be 30\% of the statistical error.
The BAO fits in configuration and Fourier space also both perform well, and 
we take conservative systematic errors of $\sigma_{\apar}=0.010$ and $\sigma_{\aperp}=0.007$.

In addition to calculating a systematic error for the individual models, the results using
the \Neveux\ and \Hou\ models are combined into a single consensus result. The results
from the two models are combined using the method of \citet{Sanchez2017a}.
Despite the results of the two analyses being highly correlated, by combining
the results, further constraints are placed on the allowed parameter space, resulting
in the combined results having a smaller uncertainty than each individual analysis.
The systematic error in the consensus result is shown in Table.~\ref{tab:systematic_errors}.
The systematic errors of the combined consensus result are reduced to $\sigma_{\fsig}=0.011$, 
$\sigma_{\apar}=0.008$ and $\sigma_{\aperp}=0.005$.
These systematic errors are included as part of the quoted errors in the final eBOSS
DR16 consensus results.

\begin{table} 
\caption{Final systematic errors in $\fsig$, $\apar$ and $\aperp$, from the \Hou\
and \Neveux\ models, combined consensus, and BAO fits.}
\begin{tabular}{cccc}
\hline
& $\fsig$ & $\apar$ & $\aperp$ \\
\hline
\Neveux             & 0.0123 & 0.0098 & 0.0066 \\
\Hou                & 0.0131 & 0.0117 & 0.0078 \\
Consensus           & 0.0106 & 0.0079 & 0.0048 \\
BAO (Fourier)       & -      & 0.0098 & 0.0055 \\
BAO (Configuration) & -      & 0.0102 & 0.0067 \\
\hline
\end{tabular}
\label{tab:systematic_errors}
\end{table}

\section{Conclusions} \label{sec:conclusions}

The eBOSS quasar sample, in the redshift range $0.8<z<2.2$, can be used as a direct 
tracer of the matter density field, enabling measurements to be taken of
the growth rate, Hubble distance and transverse comoving distance at this redshift.
These measurements can be used to place constraints on theories of dark energy,
and models of modified gravity.
The aim of the quasar mock challenge is to validate the RSD and BAO models used in
in the analysis of \citet{Hou2020} and \citet{Neveux2020}, in configuration 
and Fourier space, and estimate the systematic uncertainties.

Mock catalogues are generated using the OuterRim N-body simulation, which has a box size
of $3~\hGpc$, and was run in a WMAP7 cosmology. The box is populated with quasars using
20 different HODs, which are tuned to give approximately the same clustering measurements
as the eBOSS DR16 quasar sample,
and the same number density. We populate the snapshot at $z=1.433$, which is closest
to the effective redshift of the data, $\zeff=1.48$.

For the non-blind part of the mock challenge, 100 individual mocks are generated from the OuterRim box
for each of the 20 HODs,
which are analysed using the OuterRim cosmology as the fiducial cosmology. For each mock, we
create a version with no redshift smearing, with Gaussian redshift smearing, and a more realistic
double Gaussian redshift smearing. We also create a set of mocks that contains realistic 
smearing and 1.5\% catastrophic redshifts, which are generated from a uniform redshift distribution.

For the non-blind mocks with no smearing, the \Neveux, \Hou\ and CLPT models all perform
equally well, and there is no reason to choose any particular model over any other. 
In the Fourier space analysis of \citet{Neveux2020}, the \Neveux\ model is used, while \Hou\ is
used for the configuration space analysis of \citet{Hou2020}. 
These models are able to measure the parameters
$\fsig$ within 3\%, and $\apar$ and $\aperp$ to within 1\%, with and without redshift smearing. 
The inclusion of catastrophic redshifts
dilutes the clustering signal, leading to the measured $\fsig$ being biased towards smaller 
values, but this is still within the limit of $\sim 3\%$.

Blind mocks are generated from the OuterRim simulation, using the method of \citet{Mead2014a}
to rescale the halo positions, velocities and masses, in order to mimic a simulation that was 
run in a different target cosmology. We rescale the OuterRim simulation to 8 new cosmologies, and
generate 8 sets of 100 mocks, using the same HOD for each. The CLPT model is used to validate
the rescaled cosmologies, by recovering the expected values of $\fsig$, $\apar$ and $\aperp$. 
When the OuterRim cosmology is used as the fiducial cosmology, the scatter in the
measured values of the parameters is larger for all models, due to the incorrect assumption
of the fiducial cosmology. While the BAO peak position is scaled by the expected $\alpha$ parameters, 
the incorrect assumption of the fiducial cosmology distorts the clustering on small scales, leading
to clustering measurements which do not match the shape of the template. This results in a biased
measurement of $\fsig$, $\apar$ and $\aperp$, which is correlated with $\Omega_\mathrm{m}$.
We also generate a set of 24 blind mocks, using the same rescaled OuterRim snapshots, but 
with the HOD varied, and an additional scaling applied to velocities.

To calculate the systematic error, we combine the results from the 24 blind mocks, with
the 20 non-blind mocks with catastrophic redshifts, to take into account the effects of
the incorrect assumption of the fiducial cosmology, different HODs, redshift smearing and
catastrophic redshifts. We calculate the rms of the blind and non-blind mocks, which are
added in quadrature to estimate the total error.
We use the non-blind mocks with no smearing to correct the results from the \Hou\ model for 
having the parameter $\sigma_\mathrm{zerr}$ free, which biases the results when the mocks 
do not contain redshift smearing.

The \Neveux\ and \Hou\ models perform equally well, and we take systematic errors
for both models of $\sigma_{\fsig}=0.013$, $\sigma_{\apar}=0.012$ and $\sigma_{\aperp}=0.008$.
For the BAO fits we take systematic errors of $\sigma_{\apar}=0.010$ and $\sigma_{\aperp}=0.007$.
These results are a conservative upper bound, since they assume the errors in the blind and
non-blind mocks are uncorrelated, and that the blind mocks cover a wide range of cosmological parameters.
The results of the \Neveux\ and \Hou\ models are combined into a single consensus result, which tightens
the constraints on the parameters, compared to the individual analyses. The
consensus systematic errors are $\sigma_{\fsig}=0.011$, $\sigma_{\apar}=0.08$ and $\sigma_{\aperp}=0.005$.
These errors are combined with the errors in the final consensus results from the analysis of the
eBOSS DR16 quasar sample.
\highlight{The tests that we have performed in this work show that the measurements of the BAO
scale is very robust, despite the effects of redshift uncertainty, and the wide range of HOD models 
we have used. We can therefore be confident that using such methods will enable accurate 
cosmological distance measurements in future surveys.}

\highlight{The mock challenge presented in this paper is similar to what has been done previously in the
BOSS survey. Future surveys, such as DESI, LSST and Euclid, will cover much larger volumes, 
and aim to very precise, sub-percent level cosmological measurements. In order to achieve this precision, 
it will be essential that systematics in the models can be reduced as much as possible.
Using mock catalogues based on a single N-body simulation, as was done in this work, will not be 
good enough, and assessing the systematics will require large numbers of N-body simulations, 
with different cosmological parameters, both within \LCDM, and also extending to models beyond \LCDM. 
For the DESI survey, many of the simulations and mock catalogues required have already been produced, 
and work on the mock challenge is currently under way.}

\section*{Acknowledgements}

AS would like to thank Alexander Mead, John Peacock and Shaun Cole for helpful discussions.
AS acknowledges support from grant ANR-16-CE31-0021, eBOSS and 
ANR-17-CE31-0024-01, NILAC.
SA is supported by the European Research Council through the COSFORM Research Grant (\#670193).
GR acknowledges support from the National Research Foundation of Korea (NRF) through Grants No. 2017R1E1A1A01077508 and No. 2020R1A2C1005655 funded by the Korean Ministry of Education, Science and Technology (MoEST), and from the faculty research fund of Sejong University.

Funding for the Sloan Digital Sky Survey IV has been provided by the Alfred P. Sloan Foundation, the U.S. Department of Energy Office of Science, and the Participating Institutions. SDSS-IV acknowledges
support and resources from the Center for High-Performance Computing at
the University of Utah. The SDSS web site is www.sdss.org.
In addition, this research relied on resources provided to the eBOSS
Collaboration by the National Energy Research Scientific Computing
Center (NERSC).  NERSC is a U.S. Department of Energy Office of Science
User Facility operated under Contract No. DE-AC02-05CH11231.

SDSS-IV is managed by the Astrophysical Research Consortium for the 
Participating Institutions of the SDSS Collaboration including the 
Brazilian Participation Group, the Carnegie Institution for Science, 
Carnegie Mellon University, the Chilean Participation Group, the French Participation Group, Harvard-Smithsonian Center for Astrophysics, 
Instituto de Astrof\'isica de Canarias, The Johns Hopkins University, Kavli Institute for the Physics and Mathematics of the Universe (IPMU) / 
University of Tokyo, the Korean Participation Group, Lawrence Berkeley National Laboratory, 
Leibniz Institut f\"ur Astrophysik Potsdam (AIP),  
Max-Planck-Institut f\"ur Astronomie (MPIA Heidelberg), 
Max-Planck-Institut f\"ur Astrophysik (MPA Garching), 
Max-Planck-Institut f\"ur Extraterrestrische Physik (MPE), 
National Astronomical Observatories of China, New Mexico State University, 
New York University, University of Notre Dame, 
Observat\'ario Nacional / MCTI, The Ohio State University, 
Pennsylvania State University, Shanghai Astronomical Observatory, 
United Kingdom Participation Group,
Universidad Nacional Aut\'onoma de M\'exico, University of Arizona, 
University of Colorado Boulder, University of Oxford, University of Portsmouth, 
University of Utah, University of Virginia, University of Washington, University of Wisconsin, 
Vanderbilt University, and Yale University.

This work used the DiRAC@Durham facility managed by the Institute for Computational Cosmology on behalf of the STFC DiRAC HPC Facility (www.dirac.ac.uk). The equipment was funded by BEIS capital funding via STFC capital grants ST/K00042X/1, ST/P002293/1, ST/R002371/1 and ST/S002502/1, Durham University and STFC operations grant ST/R000832/1. DiRAC is part of the National e-Infrastructure.

\section*{Data availability}

The non-blind mock catalogues and their clustering measurements underlying this article will be shared on reasonable request to the corresponding author. The blind mocks and their clustering measurements will be shared on reasonable request to the corresponding author with permission of Katrin Heitmann and Salman Habib.




\bibliographystyle{mnras}
\bibliography{ref} 

\bsp	
\label{lastpage}
\end{document}